\documentclass[journal,draftclsnofoot,12pt,onecolumn]{IEEEtran}
\usepackage{graphicx}
\usepackage{amssymb}
\usepackage{float}
\usepackage{color}
\usepackage{subfigure}
\usepackage{caption}
\renewcommand{\theequation}{\arabic{section}.\arabic{equation}}
\newtheorem{theorem}{Theorem}

\newtheorem{lemma}{Lemma}

\newtheorem{corollary}{Corollary}

\begin{document}

\title{Self-Secure Capacity-Achieving Feedback Schemes of Gaussian Multiple-Access Wiretap Channels with Degraded Message Sets}
\author{Bin~Dai,
        Chong~Li,~\IEEEmembership{Member,~IEEE},
        Yingbin~Liang,~\IEEEmembership{Senior Member,~IEEE},
        Zheng~Ma,~\IEEEmembership{Member,~IEEE},
        and~Shlomo~Shamai (Shitz),~\IEEEmembership{Life Fellow,~IEEE}

\thanks{B. Dai and Z. Ma are with the
School of Information Science and Technology,
Southwest JiaoTong University, Chengdu 610031, China.
e-mail: daibin@home.swjtu.edu.cn, zma@home.swjtu.edu.cn.}
\thanks{C. Li is with the Nakamoto \& Turing Labs, New York, 10018, USA,
e-mail: chongl@ntlabs.io.}
\thanks{Y. Liang is with the Department of
Electrical and Computer Engineering, The Ohio State University, Columbus, 43220, USA,
e-mail: liang.889@osu.edu.}
\thanks{S. Shamai is with the Department of Electrical Engineering, Technion-Israel Institute of Technology, Technion City, 32000, Israel,
e-mail: sshlomo@ee.technion.ac.il.}
\thanks{
This work was presented in part at the IEEE Information Theory Workshop (ITW) in April, 2021.
}
}
\maketitle

\begin{abstract}

The Schalkwijk-Kailath (SK) scheme, which achieves the capacity of the point-to-point white Gaussian channel with feedback, is secure by itself
and also achieves the secrecy capacity of the Gaussian wiretap channel with feedback, i.e., the SK scheme is a 
self-secure capacity-achieving (SSCA) feedback scheme for the Gaussian wiretap channel. For the multi-user wiretap channels, recently, it has been shown that 
Ozarow's capacity-achieving feedback scheme for the two-user Gaussian multiple-access channel (GMAC) is the 
SSCA feedback scheme for the two-user Gaussian multiple-access wiretap channel (GMAC-WT).
In this paper, first, we propose a capacity-achieving feedback scheme for the two-user GMAC with degraded message sets (GMAC-DMS),
and show that this scheme is the SSCA feedback scheme for the two-user GMAC-WT with degraded message sets (GMAC-WT-DMS).
Next, 
we extend the above scheme to the
two-user GMAC-DMS with noncausal channel state information at the transmitters (NCSIT), and show that the extended scheme is 
capacity-achieving and also a SSCA feedback scheme for the two-user GMAC-WT-DMS with NCSIT.
Finally, we derive outer bounds on the secrecy capacity regions of the two-user GMAC-WT-DMS
with or without NCSIT, and numerical results show the rate gains by the feedback.

\end{abstract}


\section{Introduction \label{secI}}
\setcounter{equation}{0}

The multiple-access channel (MAC), which characterizes the up-link of wireless communication,
has received extensive attention in the literature. The capacity regions of MAC and Gaussian MAC (GMAC) were determined
by \cite{mac2} and \cite{gmac1}, respectively.
Unlike the well known fact that feedback does not increase the capacity of a discrete memoryless channel, \cite{cover2}-\cite{cover4} found that feedback increases the capacity region of the MAC
by proposing inner bounds on the capacity region of the MAC with feedback. 
The capacity region of the MAC with feedback remains open, and it is only determined for some special cases:
\begin{itemize}

\item For the two-user GMAC with feedback, Ozarow \cite{cover7} proposed a hybrid scheme which combines the cooperative scheme in \cite{cover2} 
and the Schalkwijk-Kailath (SK) scheme \cite{sk} for the point-to-point
Gaussian channel with feedback, and showed that this scheme is capacity-achieving 
\footnote{Here note that for the $N$-user ($N\geq 3$) GMAC with feedback, the capacity region remains open.}. Subsequently, \cite{cover4.x} investigated the two-user GMAC with feedback
and noncausal channel state information at the transmitters (NCSIT), and showed that a variation of Ozarow's scheme \cite{cover7} is capacity-achieving. 

\item For the two-user MAC with degraded message sets (DMS), where two independent messages
are sent from two sources to a common destination,
the uninformed encoder only has access to one message, while
the informed encoder has access to both messages.
Though it has already been shown that
feedback does not increase the capacity region of the MAC with DMS (MAC-DMS) \cite{dms1}, 
\cite{dms2} proposed a capacity-achieving scheme for the MAC-DMS with feedback, which is 
an extension of the posterior matching scheme for the point-to-point
discrete memoryless channel with feedback \cite{pm}.

\end{itemize}

The physical layer security (PLS), which captures the fundamental limit of secure transmission over communication channels, was first investigated by Wyner
in his landmark paper on the wiretap channel (WTC) \cite{Wy}.
The secrecy capacities (channel capacities with perfect secrecy constraint)
of the discrete memoryless WTC (DM-WTC) and the Gaussian WTC (G-WTC) was determined in \cite{Wy}-\cite{CK} and \cite{hellman}, respectively.
In recent years, the PLS in multiple-access channels receives much attention. Specifically, \cite{mactifs1} studied the two-user 
Gaussian multiple-access wiretap channel (GMAC-WT), and proposed an inner bound on the secrecy capacity region.
\cite{mactifs2} investigated arbitrarily varying MAC with strong secrecy constraint, and provided bounds on its secrecy capacity region. 
\cite{mactifs3}-\cite{mactifs5} studied variations of the MAC with secrecy constraint, and proposed bounds on the corresponding secrecy capacity regions.
\cite{mactifs6} proposed cooperative jamming schemes for the multiple-access wiretap channel (MAC-WT), which enhance the secrecy capacity region. 
\cite{mactifs7} studied the MAC-WT with NCSIT,
and provided bounds on its secrecy capacity region. \cite{mactifs8} investigated the effect of feedback delay on the secrecy capacity
of the finite state MAC-WT. \cite{mactifs9} studied the secure relay schemes for the MAC-WT.

Channel feedback has been proved to be a useful tool to enhance the PLS in communication systems.
Recently, \cite{gunx} showed that the secrecy capacity of the G-WTC with feedback equals the capacity of the same model without secrecy constraint,
and it is achieved by the classical SK scheme \cite{sk} which is not designed with the consideration of secrecy, i.e., 
the SK scheme is a self-secure capacity-achieving (SSCA) feedback scheme for the G-WTC.
Based on the surprising finding of \cite{gunx}, \cite{lichong} and \cite{daitit} respectively showed that variations of the classical SK scheme
are also SSCA feedback schemes for the colored G-WTC and the G-WTC with NCSIT.
Very recently, \cite{daiisita} showed that Ozarow's scheme \cite{cover7} and its variation \cite{cover4.x} are also 
SSCA feedback schemes for the two-user GMAC-WT with or without NCSIT.

Although the SSCA feedback schemes have been well studied in the Gaussian wiretap channels and the Gaussian multiple-access wiretap channels,
such a topic remains open for the multiple-access wiretap channels with DMS \footnote{
In \cite{dms2}, a capacity-achieving feedback scheme is proposed for the MAC-DMS with feedback, but whether this scheme is self-secure or not remains unknown.}.
In this paper, we focus on the two-user GMAC-WT with DMS, and with or without NCSIT, and study how to design SSCA feedback schemes for these models.
We summarize our contribution as follows.

\textbf{1}) Since Ozarow's scheme is a SSCA feedback scheme for the two-user GMAC-WT \cite{daiisita}, it is natural to ask: is this kind of scheme also be a SSCA 
feedback scheme
for the two-user GMAC-WT with DMS (GMAC-WT-DMS)? Unfortunately, we find that though Ozarow's scheme is secure by itself, it \emph{cannot} 
achieve the capacity region
of the two-user GMAC with DMS (GMAC-DMS) and feedback, hence it is not a SSCA feedback scheme for the two-user GMAC-WT-DMS. 
In this paper, we propose a novel two-step SK-type feedback scheme
for the two-user GMAC-DMS, and show that this new feedback scheme is SSCA for the two-user GMAC-WT-DMS. The novelty of this new scheme is explained below.

In the two-user GMAC-DMS with feedback, since the informed encoder has access to both messages, we split this encoder into two parts, where one part
encodes the message with rate $R_{2}$ as the codeword $V^{N}$, and the other part together with the uninformed encoder 
encode the message with rate $R_{1}$ as the codewords $U^{N}$ and $X_{1}^{N}$. For the receiver, $U^{N}$ and $X_{1}^{N}$ are decoded first, and after
successfully decoding $U^{N}$ and $X_{1}^{N}$, the receiver subtracts them from his/her received signal and further decodes $V^{N}$.

Since $U^{N}$ and $X_{1}^{N}$ are known by the informed encoder, they can be perfectly canceled when the informed encoder encodes $V^{N}$,
i.e., the noise of the equivalent channel for $V^{N}$ is the original white Gaussian channel noise
$\eta_{1}^{N}$ of the GMAC. Hence we directly apply the classical SK scheme \cite{sk} for the point-to-point white Gaussian channel with feedback
to $V^{N}$, and from \cite{gunx}, we know that the coding scheme of $V^{N}$ is SSCA. 
However, different from the encoding scheme of $V^{N}$, since $V^{N}$ is not known by the uninformed encoder,
for the encoding scheme of $U^{N}$ and $X_{1}^{N}$, the noise of their equivalent channel is $V^{N}+\eta_{1}^{N}$,
which is \emph{non-white} Gaussian noise due to the reason that $V^{N}$ is generated by classical SK scheme \cite{sk}
and it is not independent identically distributed (i.i.d.) generated. In general, it is difficult to design a SSCA SK-type scheme for the 
\emph{non-white} Gaussian channel. However, by letting the encoder of $V^{N}$ work first (starting from time $1$), 
and the encoder of $U^{N}$ and $X_{1}^{N}$ work later (starting from time $2$), we find that the SK-type scheme of $U^{N}$ and $X_{1}^{N}$
is also SSCA, and the key step to the corresponding proof is Lemma \ref{L1} in Section \ref{secIII}, i.e., for time instant $3\leq k\leq N$,
$E[\epsilon^{'}_{k-1}\eta^{'}_{1,k}]=0$,
where $\eta^{'}_{1,k}=\eta_{1,k}+V_{k}$, $\eta_{1,k}$ and $V_{k}$ are the $k$-th components of $\eta_{1}^{N}$ and $V^{N}$, respectively,
and $\epsilon^{'}_{k-1}$ is a deterministic function of $U_{k}$ and $X_{1,k}$, which are the $k$-th components of $U^{N}$ and $X_{1}^{N}$, respectively.
Here note that Lemma \ref{L1} is surprising and novel since both $V_{k}$ and $\epsilon^{'}_{k-1}$
depend on the previous noises $\eta_{1,1},...,\eta_{1,k-1}$. By using this surprising property in Lemma \ref{L1}, we show that the 
two-step SK-type feedback scheme is SSCA for the two-user GMAC-WT-DMS.

\textbf{2}) We extend the above new feedback scheme to the two-user GMAC with NCSIT and DMS (GMAC-NCSIT-DMS), and show that this extended feedback scheme 
is also SSCA
for the two-user GMAC-WT with NCSIT and DMS (GMAC-WT-NCSIT-DMS).
The novelty of this new scheme is explained below.

In the previous two-step SK-type scheme for the two-user GMAC-DMS with feedback, after decoding one message $W_{1}$, the receiver knows $U^{N}$ and $X_{1}^{N}$.
Hence in the decoding of the other message $W_{2}$, the receiver directly subtracts $U^{N}$ and $X_{1}^{N}$ from his/her received signal 
and does a similar SK-type decoding
to obtain $W_{2}$.
However, in the two-user GMAC-NCSIT-DMS with feedback, since
the receiver does not know the state interference, after decoding $W_{1}$, the receiver \emph{cannot} obtain $U^{N}$ and $X_{1}^{N}$, 
which leads to the failure of subtracting $U^{N}$ and $X_{1}^{N}$ from the receiver's received signal. Fortunately, we find that 
after decoding $W_{1}$, though the receiver only obtains partial information about $U^{N}$ and $X_{1}^{N}$, by introducing proper offsets
into the construction of $V^{N}$, $U^{N}$ and $X_{1}^{N}$, the receiver's final estimations of the transmitted messages are the same as those in
the previous two-step SK-type scheme for the GMAC-DMS with feedback, which indicates that this modified scheme is also a SSCA feedback scheme
for the two-user GMAC-WT-NCSIT-DMS.

\textbf{3}) Outer bounds on the {\em secrecy} capacity regions of the GMAC-WT-DMS and the GMAC-WT-NCSIT-DMS
are given, and numerical results
show the rate gains by the feedback.

Throughout this paper, a random variable (RV) is denoted by an upper case letter (e.g., $X$), its value is
denoted by an lower case letter (e.g., $x$), the finite alphabet of the RV is denoted by calligraphic letter (e.g., $\mathcal{X}$),
and the probability distribution of an event $\{X=x\}$ is denoted by $P_{X}(x)$.
Random vectors and their values are denoted by a similar convention.
For example,
$X^{N}$ represents a $N$-dimensional random vector $(X_{1},...,X_{N})$,
and $x^{N}=(x_{1},...,x_{N})$ represents a vector value in $\mathcal{X}^{N}$
(the $N$-th Cartesian power of the finite alphabet $\mathcal{X}$).
In addition, define $A_{j}^{N}=(A_{j,1},A_{j,2},...,A_{j,N})$ and $a_{j}^{N}=(a_{j,1},a_{j,2},...,a_{j,N})$.
Finally,
throughout this paper, the base of the $\log$ function is $2$.

The remainder of this paper is organized as follows.
Formal definitions of the models studied in this paper and preliminary are given in Section \ref{secII}.
The SSCA feedback scheme for the GMAC-WT-DMS is given in Section \ref{secIII}.
The SSCA feedback scheme for the GMAC-WT-NCSIT-DMS is given in Section \ref{secIV}.
Section \ref{secVI} includes the summary of all results in this paper and discusses future work.

\section{Model Formulation and Preliminary}\label{secII}
\setcounter{equation}{0}

\subsection{Model Formulation-type I: the GMAC-DMS with or without feedback and secrecy constraint}\label{secII-y1}

\begin{figure}[htb]
\centering
\includegraphics[scale=0.7]{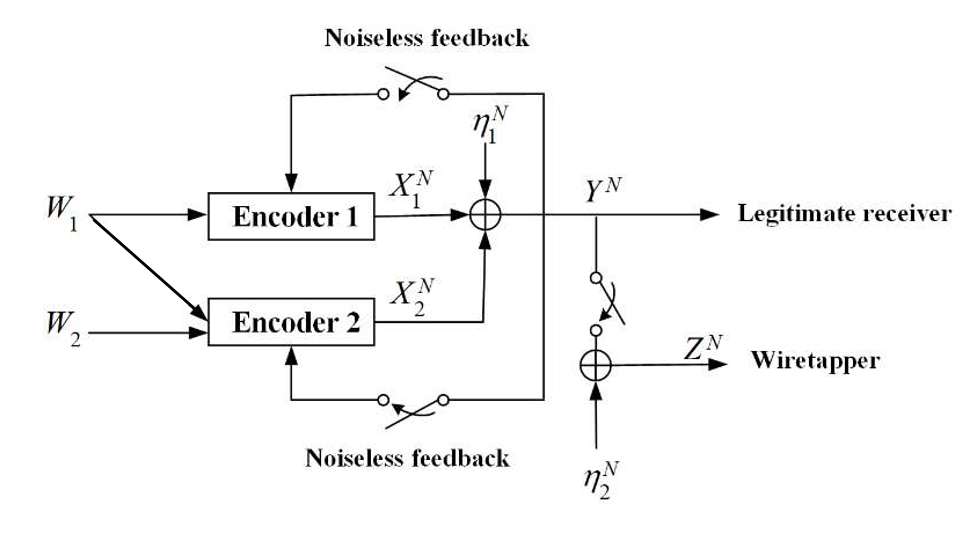}
\caption{The GMAC-DMS with or without feedback and secrecy constraint}
\label{f1}
\end{figure}

\subsubsection{Model I: The GMAC-DMS with or without feedback}\label{secII-y1-yy1}

For the GMAC-DMS with or without feedback, the $i$-th ($i\in\{1,2,...,N\}$) channel input-output relationship is given by
\begin{eqnarray}\label{wawa1}
&&Y_{i}=X_{1,i}+X_{2,i}+\eta_{1,i},
\end{eqnarray}
where $X_{1,i}$ and $X_{2,i}$ are the channel inputs subject to average power constraints $P_{1}$ and $P_{2}$ (i.e.,
$\frac{1}{N}\sum_{i=1}^{N}E[X^{2}_{1,i}]\leq P_{1}$, $\frac{1}{N}\sum_{i=1}^{N}E[X^{2}_{2,i}]\leq P_{2}$), respectively,
$Y_{i}$ is the channel output of the receiver,
and $\eta_{1,i}\sim \mathcal{N}(0, \sigma_{1}^{2})$ are the channel noises and are
i.i.d.
across the time index $i$. The message $W_{j}$ ($j=1,2$) is uniformly distributed in $\mathcal{W}_{j}=\{1,2,...,|\mathcal{W}_{j}|\}$.
For the GMAC-DMS \emph{without} feedback,
the channel input $X_{1,i}$ is a function of the message $W_{1}$,
and the channel input $X_{2,i}$ is a function of the messages $W_{1}$ and $W_{2}$. For the GMAC-DMS with feedback,
$X_{1,i}$ is a function of the message $W_{1}$ and the feedback $Y^{i-1}$,
and $X_{2,i}$ is a function of the messages $W_{1}$, $W_{2}$ and the feedback $Y^{i-1}$.
The receiver generates an estimation
$(\hat{W}_{1},\hat{W}_{2})=\psi(Y^{N})$, where $\psi$ is the legitimate receiver's decoding function, and the average decoding error probability equals
\begin{equation}\label{e204}
P_{e}=\frac{1}{|\mathcal{W}_{1}|\cdot|\mathcal{W}_{2}|}\sum_{
w_{1}\in \mathcal{W}_{1},
w_{2}\in \mathcal{W}_{2}}Pr\{\psi(y^{N})\neq (w_{1},w_{2})|(w_{1},w_{2})\,\,\mbox{sent}\}.
\end{equation}
A rate pair $(R_{1},R_{2})$ is said to be achievable if for any $\epsilon$ and sufficiently large $N$,
there exists channel encoders and decoder such that
\begin{eqnarray}\label{wawa2}
&&\frac{\log |\mathcal{W}_{1}|}{N}=R_{1},\,\,\frac{\log |\mathcal{W}_{2}|}{N}=R_{2},\,\,P_{e}\leq \epsilon.
\end{eqnarray}
The capacity regions of the GMAC-DMS with or without feedback are composed of all such achievable rate pairs, and they are
denoted by $\mathcal{C}_{gmac-dms}^{f}$ and $\mathcal{C}_{gmac-dms}$, respectively.

\subsubsection{Model II: The GMAC-WT-DMS with or without feedback}\label{secII-y1-yy3}

For the GMAC-WT-DMS with or without feedback, the $i$-th ($i\in\{1,2,...,N\}$) channel input-output relationships are given by
\begin{eqnarray}\label{wawa1.r1}
&&Y_{i}=X_{1,i}+X_{2,i}+\eta_{1,i},\,\,Z_{i}=Y_{i}+\eta_{2,i},
\end{eqnarray}
where $X_{1,i}$ and $X_{2,i}$ are the channel inputs subject to average power constraints $P_{1}$ and $P_{2}$, respectively,
$Y_{i}$ and $Z_{i}$ are the channel outputs of the legitimate receiver and the wiretapper, respectively,
and $\eta_{1,i}\sim \mathcal{N}(0, \sigma_{1}^{2})$, $\eta_{2,i}\sim \mathcal{N}(0, \sigma_{2}^{2})$ are the channel noises and are
independent identically distributed (i.i.d.)
across the time index $i$. The message $W_{j}$ ($j=1,2$) is uniformly distributed in $\mathcal{W}_{j}=\{1,2,...,|\mathcal{W}_{j}|\}$.
For the GMAC-WT-DMS \emph{without} feedback,
the channel input $X_{1,i}$ is a (stochastic) function of the message $W_{1}$,
and the channel input $X_{2,i}$ is a (stochastic) function of the messages $W_{1}$ and $W_{2}$. For the GMAC-WT-DMS with feedback,
$X_{1,i}$ is a (stochastic) function of the message $W_{1}$ and the feedback $Y^{i-1}$,
and $X_{2,i}$ is a (stochastic) function of the messages $W_{1}$, $W_{2}$ and the feedback $Y^{i-1}$.
The legitimate receiver generates an estimation
$(\hat{W}_{1},\hat{W}_{2})=\psi(Y^{N})$, where $\psi$ is the legitimate receiver's decoding function, and the average decoding error probability $P_{e}$
is defined the same as that in (\ref{e204}).
The wiretapper's equivocation rate of the messages $W_{1}$ and $W_{2}$ is defined as
\begin{equation}\label{e205.ff}
\Delta=\frac{1}{N}H(W_{1},W_{2}|Z^{N}).
\end{equation}
A rate pair $(R_{1},R_{2})$ is said to be achievable with perfect weak secrecy if for any $\epsilon$ and sufficiently large $N$,
there exists channel encoders and decoder such that
\begin{eqnarray}\label{wawa2}
&&\frac{\log |\mathcal{W}_{1}|}{N}=R_{1},\,\,\frac{\log |\mathcal{W}_{2}|}{N}=R_{2},\,\,\Delta\geq R_{1}+R_{2}-\epsilon,\,\,P_{e}\leq \epsilon.
\end{eqnarray}
The {\em secrecy} capacity regions of the GMAC-WT-DMS with or without feedback are composed of all such achievable rate pairs, 
and they are
denoted by $\mathcal{C}_{s,gmac-dms}^{f}$ and $\mathcal{C}_{s,gmac-dms}$, respectively.

\subsection{Model Formulation-type II: The GMAC-NCSIT-DMS
with or without feedback and secrecy constraint}\label{secII-y2}

\begin{figure}[htb]
\centering
\includegraphics[scale=0.7]{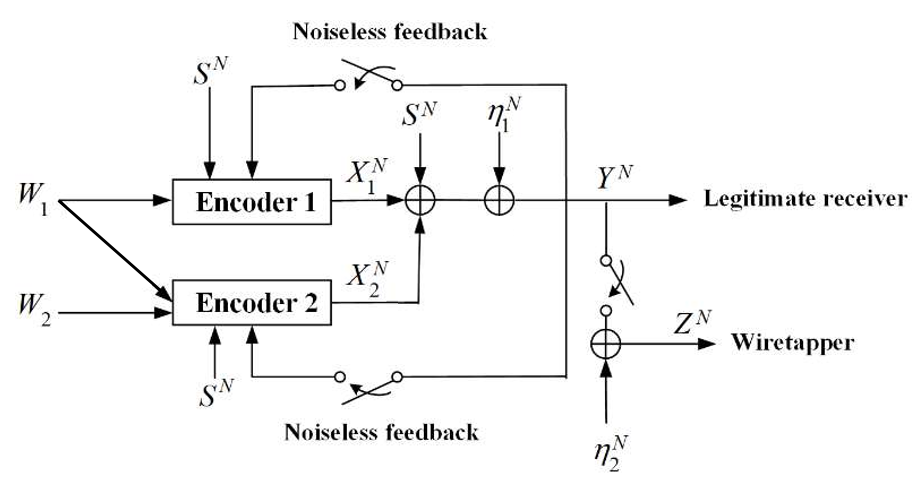}
\caption{The GMAC-NCSIT-DMS
with or without feedback and secrecy constraint}
\label{f2}
\end{figure}

\subsubsection{Model III: The GMAC-NCSIT-DMS with or without feedback}\label{secII-y2-yy1}

For the GMAC-NCSIT-DMS with or without feedback, at each time $i$ ($i\in\{1,2,...,N\}$), the channel input-output relationships are given by
\begin{eqnarray}\label{ass1}
&&Y_{i}=X_{1,i}+X_{2,i}+S_{i}+\eta_{1,i},
\end{eqnarray}
where $X_{1,i}$, $X_{2,i}$, $\eta_{1,i}$ and $Y_{i}$ are defined in the same fashion as those in Subsubsection \ref{secII-y1-yy1},
and $S_{i}\sim \mathcal{N}(0, Q)$ is the independent Gaussian state interference and is i.i.d. across the
time index $i$. The message $W_{j}$ ($j=1,2$) is uniformly distributed in $\mathcal{W}_{j}=\{1,2,...,|\mathcal{W}_{j}|\}$.
For the GMAC-NCSIT-DMS \emph{without} feedback,
the channel input $X_{1,i}$ is a function of the message $W_{1}$ and the state interference $S^{N}$,
and the channel input $X_{2,i}$ is a function of the messages $W_{1}$, $W_{2}$ and the state interference $S^{N}$. For the GMAC-NCSIT-DMS with feedback,
$X_{1,i}$ is a function of the message $W_{1}$, the state interference $S^{N}$ and the feedback $Y^{i-1}$,
and $X_{2,i}$ is a function of the messages $W_{1}$, $W_{2}$, the state interference $S^{N}$ and the feedback $Y^{i-1}$.
The receiver's decoding function, the average decoding error probability and the achievable rate pair
are defined in the same fashion as those in Subsubsection \ref{secII-y1-yy1}.
The capacity regions of the GMAC-NCSIT-DMS with or without feedback are 
denoted by $\mathcal{C}_{gmac-ncsit-dms}^{f}$ and $\mathcal{C}_{gmac-ncsit-dms}$, respectively.

\subsubsection{Model IV: The GMAC-WT-NCSIT-DMS with or without feedback}\label{secII-y2-yy3}

For the GMAC-WT-NCSIT-DMS with or without feedback, the $i$-th ($i\in\{1,2,...,N\}$) channel input-output relationships are given by
\begin{eqnarray}\label{wawa1.r}
&&Y_{i}=X_{1,i}+X_{2,i}+S_{i}+\eta_{1,i},\,\,Z_{i}=Y_{i}+\eta_{2,i},
\end{eqnarray}
where $X_{1,i}$, $X_{2,i}$, $S_{i}$, $\eta_{1,i}$ and $Y_{i}$ are defined in the same fashion as those in Subsubsection \ref{secII-y2-yy1},
and $Z_{i}$ and $\eta_{2,i}$ are defined in the same fashion as those in Subsubsection \ref{secII-y1-yy3}.
The message $W_{j}$ ($j=1,2$) is uniformly distributed in $\mathcal{W}_{j}=\{1,2,...,|\mathcal{W}_{j}|\}$.
For the GMAC-WT-NCSIT-DMS \emph{without} feedback,
$X_{1,i}$ is a (stochastic) function of the message $W_{1}$ and the state interference $S^{N}$,
and $X_{2,i}$ is a (stochastic) function of the messages $W_{1}$, $W_{2}$ and the state interference $S^{N}$.
For the GMAC-WT-NCSIT-DMS with feedback, $X_{1,i}$ is a (stochastic) function of the message $W_{1}$, the state interference $S^{N}$ and the feedback
$Y^{i-1}$, and $X_{2,i}$ is a (stochastic) function of the messages $W_{1}$, $W_{2}$, the state interference $S^{N}$ and the feedback
$Y^{i-1}$. The legitimate receiver's decoding function, the wiretapper's equivocation rate and the achievable rate pair with perfect weak secrecy are defined
in the same fashion as those in Subsubsection \ref{secII-y1-yy3}. The {\em secrecy} capacity regions of the GMAC-WT-NCSIT-DMS with or without feedback
are denoted by $\mathcal{C}_{s,gmac-ncsit-dms}^{f}$ and $\mathcal{C}_{s,gmac-ncsit-dms}$, respectively.

\subsection{Preliminary: the SK scheme for the point-to-point white Gaussian channel with feedback}\label{secII-y3-1}

For the white Gaussian channel with feedback,
at each time $i$ ($i\in\{1,2,...,N\}$), the channel input-output relationship is given by
\begin{eqnarray}\label{lss1}
&&Y_{i}=X_{i}+\eta_{1,i},
\end{eqnarray}
where $X_{i}$ is the channel input subject to an average power constraint $P$,
$Y_{i}$ is the channel output of the receiver,
and $\eta_{1,i}\sim \mathcal{N}(0, \sigma_{1}^{2})$ is the white Gaussian noise and it is
i.i.d.
across the time index $i$. The message $W$ is uniformly distributed in $\mathcal{W}=\{1,2,...,|\mathcal{W}|\}$.
The channel input $X_{i}$ is a function of the message $W$ and the feedback $Y^{i-1}$.
The receiver generates an estimation
$\hat{W}=\psi(Y^{N})$, where $\psi$ is the receiver's decoding function, and the average decoding error probability equals
\begin{equation}\label{lss1lss2}
P_{e}=\frac{1}{|\mathcal{W}|}\sum_{
w\in \mathcal{W}}Pr\{\psi(y^{N})\neq w|w\,\,\mbox{sent}\}.
\end{equation}
The capacity of the white Gaussian channel with feedback is denoted by $\mathcal{C}_{g}^{f}$, and it equals the capacity
$\mathcal{C}_{g}$ of the white Gaussian channel, which is given by
\begin{equation}\label{lss3}
\mathcal{C}_{g}^{f}=\mathcal{C}_{g}=\frac{1}{2}\log(1+\frac{P}{\sigma_{1}^{2}}).
\end{equation}
In \cite{sk}, it has been shown that SK scheme achieves $\mathcal{C}_{g}^{f}$, and this classical scheme is briefly described below.

Since $W$ takes values in $\mathcal{W}=\{1,2,...,2^{NR}\}$, we divide the interval $[-0.5,0.5]$ into
$2^{NR}$ equally spaced sub-intervals, and the center of each
sub-interval is mapped to a message value in $\mathcal{W}$. Let $\theta$ be the center of the sub-interval w.r.t. the message $W$
(the variance of $\theta$ approximately equals $\frac{1}{12}$).
At time $1$, the transmitter sends
\begin{eqnarray}\label{qe1}
&&X_{1}=\sqrt{12P}\theta.
\end{eqnarray}
The receiver obtains $Y_{1}=X_{1}+\eta_{1,1}$, and gets an estimation of $\theta$ by computing
\begin{eqnarray}\label{qe2}
&&\hat{\theta}_{1}=\frac{Y_{1}}{\sqrt{12P}}=\theta+\frac{\eta_{1,1}}{\sqrt{12P}}=\theta+\epsilon_{1},
\end{eqnarray}
where $\epsilon_{1}=\hat{\theta}_{1}-\theta=\frac{\eta_{1}}{\sqrt{12P}}$. Let $\alpha_{1}\triangleq Var(\epsilon_{1})=\frac{\sigma_{1}^{2}}{12P}$.

At time $2\leq k\leq N$, the receiver obtains $Y_{k}=X_{k}+\eta_{1,k}$, and gets an estimation of $\theta_{k}$ by computing
\begin{eqnarray}\label{qe8}
&&\hat{\theta}_{k}=\hat{\theta}_{k-1}-\frac{E[Y_{k}\epsilon_{k-1}]}{E[Y^{2}_{k}]}Y_{k},
\end{eqnarray}
where $\epsilon_{k}=\hat{\theta}_{k}-\theta$, (\ref{qe8}) yields that
\begin{eqnarray}\label{qe9}
&&\epsilon_{k}=\epsilon_{k-1}-\frac{E[Y_{k}\epsilon_{k-1}]}{E[Y^{2}_{k}]}Y_{k}.
\end{eqnarray}
Meanwhile, for time $2\leq k\leq N$, the transmitter sends
\begin{eqnarray}\label{qe5}
&&X_{k}=\sqrt{\frac{P}{\alpha_{k-1}}}\epsilon_{k-1},
\end{eqnarray}
where $\alpha_{k-1}\triangleq Var(\epsilon_{k-1})$.

In \cite{sk}, it has been shown that if $R<\frac{1}{2}\log(1+\frac{P}{\sigma_{1}^{2}})$,
$P_{e}\rightarrow 0$ as $N\rightarrow \infty$.

\section{The SSCA feedback scheme for the GMAC-WT-DMS}\label{secIII}
\setcounter{equation}{0}

In this section, first, we propose a two-step SK-type
feedback scheme that achieves the capacity of GMAC-DMS with feedback. Second, we show that the proposed feedback scheme is secure by itself and also achieves
the secrecy capacity region $\mathcal{C}^{f}_{s,gmac-dms}$ of the GMAC-WT-DMS with feedback.
Finally, in order to show the
rate gains by the feedback, an outer bound on the secrecy capacity region $\mathcal{C}_{s,gmac-dms}$ of GMAC-WT-DMS is provided,
and the capacity results given in this section are further explained via a numerical example.

\subsection{A capacity-achieving two-step SK-type scheme for the GMAC-DMS with feedback}\label{secIII-1}

The model of the GMAC-DMS with feedback is formulated in Section \ref{secII-y1-yy1}.
In this subsection, first, we introduce capacity results on GMAC-DMS with or without feedback. Then, we propose a two-step SK-type scheme and show that
this scheme achieves the capacity of GMAC-DMS with feedback.

\subsubsection{Capacity results on GMAC-DMS with or without feedback}\label{secIII-1-1}

The following Corollary \ref{bbk1} characterizes the capacity region $\mathcal{C}_{gmac-dms}$ of the GMAC-DMS.
\begin{corollary}\label{bbk1}
The capacity region $\mathcal{C}_{gmac-dms}$ of the GMAC-DMS is given by
\begin{eqnarray}\label{clss1}
\mathcal{C}_{gmac-dms}=\bigcup_{0\leq \rho\leq 1}&&\left\{(R_{1}\geq 0,R_{2}\geq 0):R_{2}\leq\frac{1}{2}\log\left(1+\frac{P_{2}(1-\rho^{2})}{\sigma_{1}^{2}}\right),\right.\nonumber\\
&&\left.R_{1}+R_{2}\leq \frac{1}{2}\log\left(1+\frac{P_{1}+P_{2}+2\sqrt{P_{1}P_{2}}\rho}{\sigma_{1}^{2}}\right)\right\}.
\end{eqnarray}
\end{corollary}

\begin{IEEEproof}

\emph{Achievability of $\mathcal{C}_{gmac-dms}$}: From \cite{mac-dms}, the capacity region $\mathcal{C}_{mac-dms}$ of
the discrete memoryless MAC-DMS (DM-MAC-DMS) is given by
\begin{eqnarray}\label{clss2}
&&\mathcal{C}_{mac-dms}=\{(R_{1},R_{2}):R_{2}\leq I(X_{2};Y|X_{1}),\,\,R_{1}+R_{2}\leq I(X_{1},X_{2};Y)\}
\end{eqnarray}
for some joint distribution $P_{X_{1}X_{2}}(x_{1},x_{2})$.
Then, substituting $X_{1}\sim \mathcal{N}(0, P_{1})$
and $X_{2}\sim \mathcal{N}(0, P_{2})$ and (\ref{wawa1}) into (\ref{clss2}), defining $\rho=\frac{E[X_{1}X_{2}]}{\sqrt{P_{1}P_{2}}}$, and
following the idea of the encoding-decoding scheme of \cite{mac-dms},
the achievability of $\mathcal{C}_{gmac-dms}$ is proved.

\emph{Converse of $\mathcal{C}_{gmac-dms}$}: the converse proof of $\mathcal{C}_{gmac-dms}$ follows the idea of
the converse part in GMAC with feedback \cite[pp. 627-628]{cover7} (see the converse proof of the bounds on $R_{2}$ and $R_{1}+R_{2}$), and hence
we omit the details here. The proof of Corollary \ref{bbk1} is completed.
\end{IEEEproof}

In \cite{dms1}, it has been shown that feedback does not increase
the capacity region $\mathcal{C}_{gmac-dms}$ of the GMAC-DMS, i.e.,
\begin{eqnarray}\label{xhgg1}
&&\mathcal{C}^{f}_{gmac-dms}=\mathcal{C}_{gmac-dms},
\end{eqnarray}
where $\mathcal{C}_{gmac-dms}$ is given in (\ref{clss1}).
Here note that though the capacity region $\mathcal{C}^{f}_{gmac-dms}$ of the GMAC-DMS with feedback is determined, the SK-type
feedback scheme that achieves $\mathcal{C}^{f}_{gmac-dms}$ remains unknown. In the remainder of this section, first, a two-step SK-type feedback scheme
is proposed for the GMAC-DMS with feedback, and it is shown to be capacity-achieving. Then, we will show that this two-step SK-type scheme also achieves
the secrecy capacity region $\mathcal{C}^{f}_{s,gmac-dms}$ of the GMAC-WT-DMS with feedback.

\subsubsection{A capacity-achieving two-step SK-type feedback scheme for the GMAC-DMS with feedback}\label{secIII-1-2}

\begin{figure}[htb]
\centering
\includegraphics[scale=0.5]{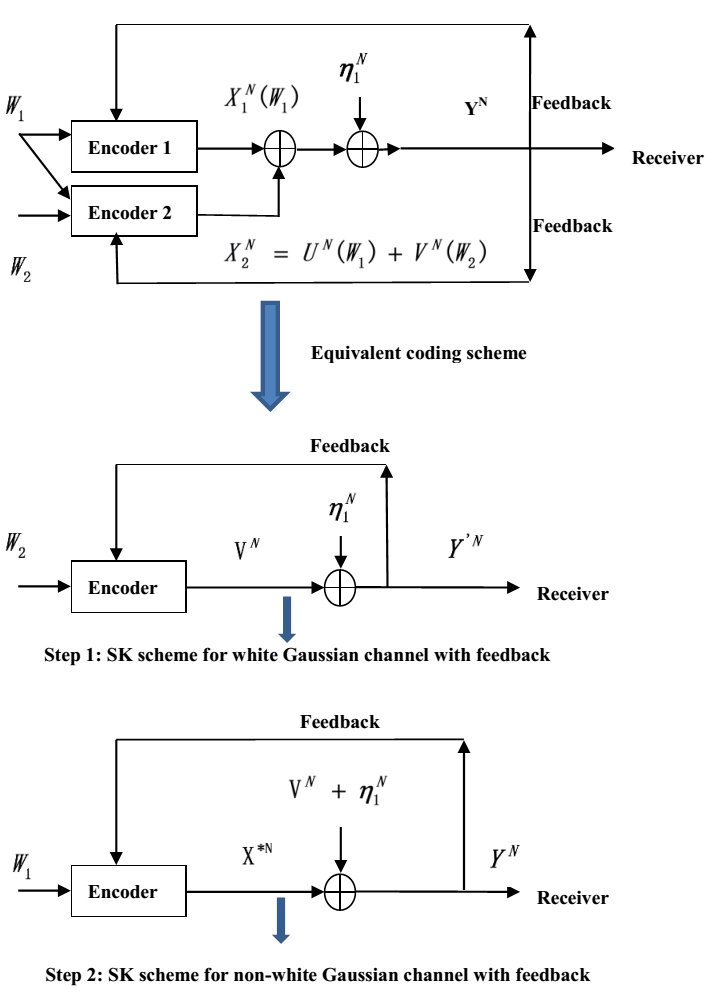}
\caption{Two-step SK-type feedback scheme for GMAC-DMS with feedback}
\label{f12}
\end{figure}

The main idea of the two-step SK-type feedback scheme is briefly illustrated by the following Figure \ref{f12}. In Figure \ref{f12},
the common message $W_{1}$ is encoded by both transmitters, and the private message $W_{2}$ is only available at Transmitter $2$.
Specifically, Transmitter $1$ uses power $P_{1}$ to encode $W_{1}$ and the feedback $Y^{N}$ as $X_{1}^{N}$. Transmitter $2$ uses power $(1-\rho^{2})P_{2}$
to encode $W_{2}$ and $Y^{N}$ as $V^{N}$, and power $\rho^{2}P_{2}$ to encode $W_{1}$ and $Y^{N}$ as $U^{N}$, where $0\leq \rho \leq 1$,
\begin{eqnarray}\label{clss3}
&&X_{2}^{N}=U^{N}+V^{N},
\end{eqnarray}
and the average transmission power of $X_{2}^{N}$ tends to $P_{2}$ for large $N$ will be explained later.
Here note that since $W_{1}$ is known by Transmitter $2$, the codewords $X_{1}^{N}$ and $U^{N}$ can be subtracted when
applying SK scheme to $W_{2}$, i.e., for the SK scheme of $W_{2}$, the equivalent channel model has input $V^{N}$, output
$Y^{'N}=Y^{N}-X_{1}^{N}-U^{N}$, and channel noise $\eta_{1}^{N}$.

In addition, since $W_{1}$ is known by both transmitters and $W_{2}$ is only available at Transmitter $2$,
for the SK scheme of $W_{1}$, the equivalent channel model has inputs $X_{1}^{N}$ and $U^{N}$,
output $Y^{N}$, and channel noise $\eta_{1}^{N}+V^{N}$, which is non-white Gaussian noise since $V^{N}$ is not i.i.d. generated. Furthermore, observing that
\begin{eqnarray}\label{clss4}
&&Y_{i}=X_{1,i}+U_{i}+V_{i}+\eta_{1,i}=X^{*}_{i}+V_{i}+\eta_{1,i},
\end{eqnarray}
where $X^{*}_{i}=X_{1,i}+U_{i}$, $X^{*}_{i}$ is Gaussian distributed with zero mean and variance $P_{i}^{*}$,
\begin{eqnarray}\label{clss4.rr1}
&&P_{i}^{*}=P_{1}+\rho^{2}P_{2}+2\sqrt{P_{1}P_{2}}\rho\rho^{'}_{i}\leq P_{1}+\rho^{2}P_{2}+2\sqrt{P_{1}P_{2}}\rho=P^{*},
\end{eqnarray}
$\rho^{'}_{i}=\frac{E[X_{1,i}U_{i}]}{\rho\sqrt{P_{1}P_{2}}}$ and $0\leq \rho^{'}_{i} \leq 1$.
Hence for the SK scheme of $W_{1}$, the input of the equivalent channel model
can be viewed as $X^{*}_{i}$. Since $X_{1,i}$ is known by Transmitter $2$, let
\begin{eqnarray}\label{clss4.rr2}
&&U_{i}=\rho\sqrt{\frac{P_{2}}{P_{1}}}X_{1,i}.
\end{eqnarray}
Then we have $\rho^{'}_{i}=1$, which leads to
\begin{eqnarray}\label{clss4.rr3}
&&P_{i}^{*}=P^{*}=P_{1}+\rho^{2}P_{2}+2\sqrt{P_{1}P_{2}}\rho,
\end{eqnarray}
and $X^{*}_{i}\sim \mathcal{N}(0,P^{*})$.
The encoding and decoding procedure of Figure \ref{f12} is described below.

Since $W_{j}$ ($j=1,2$) takes values in $\mathcal{W}_{j}=\{1,2,...,2^{NR_{j}}\}$, divide the interval $[-0.5,0.5]$ into
$2^{NR_{j}}$ equally spaced sub-intervals, and the center of each
sub-interval is mapped to a message value in $\mathcal{W}_{j}$. Let $\theta_{j}$ be the center of the sub-interval w.r.t. the message $W_{j}$
(the variance of $\theta_{j}$ approximately equals $\frac{1}{12}$).

\emph{Encoding:}
At time $1$, Transmitter $1$ sends
\begin{eqnarray}\label{sheva1}
&&X_{1,1}=0.
\end{eqnarray}
Transmitter $2$ sends
\begin{eqnarray}\label{qqe1}
&&V_{1}=\sqrt{12(1-\rho^{2})P_{2}}\theta_{2},
\end{eqnarray}
and
\begin{eqnarray}\label{sheva2}
&&U_{1}=\rho\sqrt{\frac{P_{2}}{P_{1}}}X_{1,1}=0.
\end{eqnarray}
The receiver obtains $Y_{1}=X_{1,1}+X_{2,1}+\eta_{1,1}=X_{1,1}+V_{1}+U_{1}+\eta_{1,1}=V_{1}+\eta_{1,1}$, and sends $Y_{1}$ back to Transmitter $2$.
Let $Y^{'}_{1}=Y_{1}=V_{1}+\eta_{1,1}$, Transmitter $2$ computes
\begin{eqnarray}\label{qqe2}
&&\frac{Y^{'}_{1}}{\sqrt{12(1-\rho^{2})P_{2}}}=\theta_{2}+\frac{\eta_{1,1}}{\sqrt{12(1-\rho^{2})P_{2}}}=\theta_{2}+\epsilon_{1}.
\end{eqnarray}
Let $\alpha_{1}\triangleq Var(\epsilon_{1})=\frac{\sigma_{1}^{2}}{12(1-\rho^{2})P_{2}}$.

At time $2$, Transmitter $2$ sends
\begin{eqnarray}\label{qqe1.xxs1}
&&V_{2}=\sqrt{\frac{(1-\rho^{2})P_{2}}{\alpha_{1}}}\epsilon_{1}.
\end{eqnarray}
On the other hand, at time $2$,
Transmitters $1$ and $2$ respectively send $X_{1,2}$ and $U_{2}=\rho\sqrt{\frac{P_{2}}{P_{1}}}X_{1,2}$ such that
\begin{eqnarray}\label{qqe1.r1}
&&X_{2}^{*}=U_{2}+X_{1,2}=\sqrt{12P^{*}}\theta_{1}.
\end{eqnarray}
Once receiving the feedback $Y_{2}=X_{2}^{*}+V_{2}+\eta_{1,2}$, both transmitters compute
\begin{eqnarray}\label{qqe2.r2}
&&\frac{Y_{2}}{\sqrt{12P^{*}}}=\theta_{1}+\frac{V_{2}+\eta_{1,2}}{\sqrt{12P^{*}}}=\theta_{1}+\epsilon^{'}_{2}.
\end{eqnarray}
and send $X_{1,3}$ and $U_{3}=\rho\sqrt{\frac{P_{2}}{P_{1}}}X_{1,3}$ such that
\begin{eqnarray}\label{sheva5}
&&X_{3}^{*}=U_{3}+X_{1,3}=\sqrt{\frac{P^{*}}{\alpha^{'}_{2}}}\epsilon^{'}_{2},
\end{eqnarray}
where $\alpha^{'}_{2}\triangleq Var(\epsilon^{'}_{2})$.
In addition, subtracting $X_{1,2}$ and $U_{2}$ from $Y_{2}$ and let $Y^{'}_{2}=Y_{2}-X_{1,2}-U_{2}=V_{2}+\eta_{1,2}$,
Transmitter $2$ computes
\begin{eqnarray}\label{sheva3}
&&\epsilon_{2}=\epsilon_{1}-\frac{E[Y^{'}_{2}\epsilon_{1}]}{E[(Y^{'}_{2})^{2}]}Y^{'}_{2}.
\end{eqnarray}
and sends
\begin{eqnarray}\label{sheva4}
&&V_{3}=\sqrt{\frac{(1-\rho^{2})P_{2}}{\alpha_{2}}}\epsilon_{2},
\end{eqnarray}
where $\alpha_{2}\triangleq Var(\epsilon_{2})$.

At time $4\leq k\leq N$, once receiving $Y_{k-1}=X_{1,k-1}+U_{k-1}+V_{k-1}+\eta_{1,k-1}$,
Transmitter $2$ computes
\begin{eqnarray}\label{qqe1.xxs3}
&&\epsilon_{k-1}=\epsilon_{k-2}-\frac{E[Y^{'}_{k-1}\epsilon_{k-2}]}{E[(Y^{'}_{k-1})^{2}]}Y^{'}_{k-1},
\end{eqnarray}
where
\begin{eqnarray}\label{sheva6}
&&Y^{'}_{k-1}=Y_{k-1}-X_{1,k-1}-U_{k-1},
\end{eqnarray}
and sends
\begin{eqnarray}\label{qqe1.xxs4}
&&V_{k}=\sqrt{\frac{(1-\rho^{2})P_{2}}{\alpha_{k-1}}}\epsilon_{k-1},
\end{eqnarray}
where $\alpha_{k-1}\triangleq Var(\epsilon_{k-1})$.
In the meanwhile, Transmitters $1$ and $2$ respectively send $X_{1,k}$ and $U_{k}=\rho\sqrt{\frac{P_{2}}{P_{1}}}X_{1,k}$ such that
\begin{eqnarray}\label{qqe1.xxs5}
&&X_{k}^{*}=U_{k}+X_{1,k}=\sqrt{\frac{P^{*}}{\alpha^{'}_{k-1}}}\epsilon^{'}_{k-1},
\end{eqnarray}
where
\begin{eqnarray}\label{qqe1.xxs6}
&&\epsilon^{'}_{k-1}=\epsilon^{'}_{k-2}-\frac{E[Y_{k-1}\epsilon^{'}_{k-2}]}{E[(Y_{k-1})^{2}]}Y_{k-1},
\end{eqnarray}
and $\alpha^{'}_{k-1}\triangleq Var(\epsilon^{'}_{k-1})$.

The following Lemma \ref{L1} is crucial for the analysis of the average transmission power of $X_{2}^{N}$ and the decoding error probability.

\begin{lemma}\label{L1}
For $3\leq k\leq N$,
\begin{eqnarray}\label{dlrb1}
&&E[\epsilon^{'}_{k-1}\eta^{'}_{1,k}]=0,
\end{eqnarray}
where
\begin{eqnarray}\label{qqe1.xxs11}
&&\eta^{'}_{1,k}=\eta_{1,k}+V_{k}.
\end{eqnarray}
\end{lemma}
\begin{IEEEproof}
See Appendix \ref{appenx}.
\end{IEEEproof}

\emph{Analysis of the average transmission power of $X_{2}^{N}$}:

The above Lemma \ref{L1} indicates that for $3\leq k\leq N$,
\begin{eqnarray}\label{dlrb1.lgd1}
&&E[\epsilon^{'}_{k-1}\eta^{'}_{1,k}]=E[\epsilon^{'}_{k-1}(\eta_{1,k}+V_{k})]
=E[\epsilon^{'}_{k-1}\eta_{1,k}]+E[\epsilon^{'}_{k-1}V_{k}]
\stackrel{(1)}=E[\epsilon^{'}_{k-1}V_{k}]=0,
\end{eqnarray}
where (1) follows from the fact that
$\eta_{1,k}$ is independent of $\epsilon^{'}_{k-1}$ ($\epsilon^{'}_{k-1}$ is a function of $(\eta_{1,1},...,\eta_{1,k-1})$).
Since
\begin{eqnarray}\label{dlrb1.lgd2}
&&U_{k}\stackrel{(2)}=\frac{\rho\sqrt{\frac{P_{2}}{P_{1}}}}{\rho\sqrt{\frac{P_{2}}{P_{1}}}+1}\sqrt{\frac{P^{*}}{\alpha^{'}_{k-1}}}\epsilon^{'}_{k-1},
\end{eqnarray}
where (2) follows from (\ref{qqe1.xxs5}) and $U_{k}=\rho\sqrt{\frac{P_{2}}{P_{1}}}X_{1,k}$,
substituting (\ref{dlrb1.lgd2}) into (\ref{dlrb1.lgd1}), we conclude that
\begin{eqnarray}\label{dlrb1.lgd3}
&&E[U_{k}V_{k}]=0,
\end{eqnarray}
for $3\leq k\leq N$. In addition, from (\ref{qqe1}), (\ref{sheva2}), (\ref{qqe1.xxs1}), (\ref{qqe1.r1}) and the fact that
$\theta_{1}$ is independent of $\eta_{1,1}$, we conclude that
\begin{eqnarray}\label{dlrb1.lgd4}
&&E[U_{1}V_{1}]=E[U_{2}V_{2}]=0,
\end{eqnarray}
and hence $E[U_{k}V_{k}]=0$ for $1\leq k\leq N$.

Here note that for $1\leq k\leq N$,
\begin{eqnarray}\label{dlrb1.lgd5}
&&E[X^{2}_{2,k}]=E[(U_{k}+V_{k})^{2}]\stackrel{(3)}=E[U^{2}_{k}]+E[V^{2}_{k}],
\end{eqnarray}
where (3) follows from $E[U_{k}V_{k}]=0$ for $1\leq k\leq N$. From the above encoding procedure, we conclude that
$E[X^{2}_{2,1}]=E[U^{2}_{1}]+E[V^{2}_{1}]=(1-\rho^{2})P_{2}$, and $E[X^{2}_{2,k}]=E[U^{2}_{k}]+E[V^{2}_{k}]=P_{2}$ for $2\leq k\leq N$,
which means that the average transmission power of $X_{2}^{N}$ tends to $P_{2}$ for large $N$.

\emph{Decoding:}

The receiver uses a two-step decoding scheme. First, from (\ref{qe8}), we observe that at time $k$ ($3\leq k\leq N$),
the receiver's estimation $\hat{\theta}_{1,k}$ of $\theta_{1}$ is given by
\begin{eqnarray}\label{qqe1.xxs7}
&&\hat{\theta}_{1,k}=\hat{\theta}_{1,k-1}-\frac{E[Y_{k}\epsilon^{'}_{k-1}]}{E[(Y_{k})^{2}]}Y_{k},
\end{eqnarray}
where $\epsilon^{'}_{k-1}=\hat{\theta}_{1,k-1}-\theta_{1}$ and it is computed by (\ref{qqe1.xxs6}), and
\begin{eqnarray}\label{qqe1.xxs8}
&&\hat{\theta}_{1,2}=\frac{Y_{2}}{\sqrt{12P^{*}}}=\theta_{1}+\frac{V_{2}+\eta_{1,2}}{\sqrt{12P^{*}}}=\theta_{1}+\epsilon^{'}_{2}.
\end{eqnarray}
Second, after decoding $W_{1}$ and the corresponding codewords $X_{1,k}$ and $U_{k}$ for all $1\leq k\leq N$, the receiver
subtracts $X_{1,k}$ and $U_{k}$ from $Y_{k}$, and obtains $Y^{'}_{k}=V_{k}+\eta_{1,k}$. At time $k$ ($1\leq k\leq N$),
the receiver's estimation $\hat{\theta}_{2,k}$ of $\theta_{2}$ is given by
\begin{eqnarray}\label{qqe1.xxs9}
&&\hat{\theta}_{2,k}=\hat{\theta}_{2,k-1}-\frac{E[Y^{'}_{k}\epsilon_{k-1}]}{E[(Y^{'}_{k})^{2}]}Y^{'}_{k},
\end{eqnarray}
where $\epsilon_{k-1}=\hat{\theta}_{2,k-1}-\theta_{2}$ and it is computed by (\ref{qqe1.xxs3}), and
\begin{eqnarray}\label{qqe1.xxs10}
&&\hat{\theta}_{2,1}=\frac{Y^{'}_{1}}{\sqrt{12(1-\rho^{2})P_{2}}}=\theta_{2}+\frac{\eta_{1,1}}{\sqrt{12(1-\rho^{2})P_{2}}}=\theta_{2}+\epsilon_{1}.
\end{eqnarray}

\emph{Decoding error probability analysis:}

The decoding error probability $P_{e}$ of the receiver is upper bounded by
\begin{eqnarray}\label{qqe1.b1}
&&P_{e}\leq P_{e1}+P_{e2},
\end{eqnarray}
where $P_{ej}$ ($j=1,2$) is the receiver's decoding error probability of $W_{j}$.
Observing that the transmission of $W_{2}$ is through an equivalent white Gaussian channel with power $(1-\rho^{2})P_{2}$ and
Gaussian noise variance $\sigma_{1}^{2}$,
hence from the classical SK scheme \cite{sk},
we conclude that the decoding error probability
$P_{e2}$ of $W_{2}$ tends to $0$ as $N\rightarrow \infty$ if
\begin{eqnarray}\label{ti1}
&&R_{2}<\frac{1}{2}\log(1+\frac{(1-\rho^{2})P_{2}}{\sigma_{1}^{2}}),
\end{eqnarray}
and hence we omit the derivation here. Now it remains to bound $P_{e1}$,
see the followings.

First, from (\ref{qqe1.xxs19}) and the fact that $\alpha^{'}_{k}=Var(\epsilon^{'}_{k})=E[(\epsilon^{'}_{k})^{2}]$, we have
\begin{eqnarray}\label{qqe1.xxs23}
&&\alpha^{'}_{k}\stackrel{(a)}=\alpha^{'}_{k-1}\left(\frac{\sqrt{\frac{P^{*}}{\alpha^{'}_{k-1}}}E[\epsilon^{'}_{k-1}\eta^{'}_{1,k}]
+(1-\rho^{2})P_{2}+\sigma_{1}^{2}}{P^{*}+2\sqrt{\frac{P^{*}}{\alpha^{'}_{k-1}}}E[\epsilon^{'}_{k-1}\eta^{'}_{1,k}]+(1-\rho^{2})P_{2}+\sigma_{1}^{2}}\right)^{2}\nonumber\\
&&-2\frac{(\sqrt{\frac{P^{*}}{\alpha^{'}_{k-1}}}E[\epsilon^{'}_{k-1}\eta^{'}_{1,k}]
+(1-\rho^{2})P_{2}+\sigma_{1}^{2})(\sqrt{P^{*}\cdot\alpha^{'}_{k-1}}+E[\epsilon^{'}_{k-1}\eta^{'}_{1,k}])}{(P^{*}+2\sqrt{\frac{P^{*}}{\alpha^{'}_{k-1}}}
E[\epsilon^{'}_{k-1}\eta^{'}_{1,k}]+(1-\rho^{2})P_{2}+\sigma_{1}^{2})^{2}}E[\epsilon^{'}_{k-1}\eta^{'}_{1,k}]\nonumber\\
&&+(\sigma_{1}^{2}+(1-\rho^{2})P_{2})\left(\frac{\sqrt{P^{*}\cdot\alpha^{'}_{k-1}}+E[\epsilon^{'}_{k-1}\eta^{'}_{1,k}]}{P^{*}
+2\sqrt{\frac{P^{*}}{\alpha^{'}_{k-1}}}E[\epsilon^{'}_{k-1}\eta^{'}_{1,k}]+(1-\rho^{2})P_{2}+\sigma_{1}^{2}}\right)^{2}\nonumber\\
&&\stackrel{(b)}=\frac{\alpha^{'}_{k-1}r^{2}(r^{2}+P^{*})}
{(P^{*}+r^{2})^{2}}=\frac{\alpha^{'}_{k-1}r^{2}}{P^{*}+r^{2}},
\end{eqnarray}
where (a) follows from (\ref{qqe1.xxs12}), and (b) follows from Lemma \ref{L1} and the definition in (\ref{qqe1.xxs24}).

Then, from (\ref{qqe1.xxs23}), we can conclude that
\begin{eqnarray}\label{qqe1.xxs26}
&&\sqrt{\alpha^{'}_{N}}
\stackrel{(c)}=(\frac{r}{\sqrt{r^{2}+P^{*}}})^{N-2}\sqrt{\alpha^{'}_{2}}\stackrel{(d)}=(\frac{r}{\sqrt{r^{2}+P^{*}}})^{N-2}\frac{r}{\sqrt{12P^{*}}},
\end{eqnarray}
where (c) follows from (\ref{qqe1.xxs23}), and (d) follows from $\alpha^{'}_{2}=Var(\epsilon^{'}_{2})$, (\ref{sheva.x1}) and (\ref{qqe1.xxs24}).

Finally, we bound $P_{e1}$ as follows. From $\epsilon^{'}_{N}=\hat{\theta}_{1,N}-\theta_{1}$ and the definition of $\theta_{1}$, we have
\begin{eqnarray}\label{qqe1.xxs27}
P_{e1}&\leq& Pr\left\{|\epsilon^{'}_{N}|>\frac{1}{2(|\mathcal{W}_{1}|-1)}\right\}\nonumber\\
&\stackrel{(e)}\leq& 2Q\left(\frac{1}{2\cdot2^{NR_{1}}}\cdot\frac{1}{\sqrt{\alpha^{'}_{N}}}\right)\nonumber\\
&\stackrel{(f)}=&2Q\left(\frac{1}{2}\cdot2^{-NR_{1}}(\frac{r}{\sqrt{r^{2}+P^{*}}})^{-N+2}\sqrt{\frac{12P^{*}}{r^{2}}}\right)\nonumber\\
&=&2Q\left(\frac{1}{2}\sqrt{\frac{12P^{*}}{r^{2}}}2^{-NR_{1}}(\frac{\sqrt{r^{2}+P^{*}}}{r})^{N-2}\right)\nonumber\\
&=&2Q\left(\frac{1}{2}\sqrt{\frac{12P^{*}}{r^{2}}}2^{-NR_{1}}2^{(N-2)\log\frac{\sqrt{r^{2}+P^{*}}}{r}}\right)\nonumber\\
&=&2Q\left(\frac{1}{2}\sqrt{\frac{12P^{*}}{r^{2}}}2^{-2\log\frac{\sqrt{r^{2}+P^{*}}}{r}}
2^{-N(R_{1}-\log\frac{\sqrt{r^{2}+P^{*}}}{r})}\right),
\end{eqnarray}
where (e) follows from $Q(x)$ is the tail of the unit Gaussian distribution evaluated at $x$, and (f) follows from (\ref{qqe1.xxs26})
and the fact that $Q(x)$ is decreasing while $x$ is increasing.
From (\ref{qqe1.xxs27}), we can conclude that if
\begin{eqnarray}\label{qqe1.xxs28}
&&R_{1}<\log\frac{\sqrt{r^{2}+P^{*}}}{r}=\frac{1}{2}\log(1+\frac{P^{*}}{r^{2}})\stackrel{(g)}
=\frac{1}{2}\log(1+\frac{P_{1}+\rho^{2}P_{2}+2\sqrt{P_{1}P_{2}}\rho}{(1-\rho^{2})P_{2}+\sigma_{1}^{2}}),
\end{eqnarray}
where (g) follows from (\ref{clss4.rr3}) and (\ref{qqe1.xxs24}),
$P_{e1}\rightarrow 0$ as $N\rightarrow \infty$.

Now we have shown if (\ref{ti1}) and (\ref{qqe1.xxs28}) are satisfied, the decoding error probability $P_{e}$ of the receiver
tends to $0$ as $N\rightarrow \infty$. In other words,
the rate pair $(R_{1}=\frac{1}{2}\log(1+\frac{P_{1}+\rho^{2}P_{2}+2\sqrt{P_{1}P_{2}}\rho}{(1-\rho^{2})P_{2}+\sigma_{1}^{2}}),
R_{2}=\frac{1}{2}\log(1+\frac{(1-\rho^{2})P_{2}}{\sigma_{1}^{2}}))$ is achievable for all $0\leq \rho\leq 1$, which indicates that all rate pairs $(R_{1},R_{2})$
in $\mathcal{C}^{f}_{gmac-dms}$ are achievable. Hence the proposed two-step SK-type feedback scheme achieves the capacity region
$\mathcal{C}^{f}_{gmac-dms}$ of GMAC-DMS with feedback.

\subsection{Capacity result on the GMAC-WT-DMS with feedback}\label{secIII-2}

The model of the GMAC-WT-DMS with feedback is formulated in Section \ref{secII-y1-yy3}. The following Theorem \ref{T5}
establishes that the secrecy constraint does not reduce the capacity of GMAC-DMS with feedback.

\begin{theorem}\label{T5}
$\mathcal{C}_{s,gmac-dms}^{f}=\mathcal{C}_{gmac-dms}$, where $\mathcal{C}_{s,gmac-dms}^{f}$ is the secrecy capacity region of the GMAC-WT-DMS with feedback,
and $\mathcal{C}_{gmac-dms}$ is given in Corollary \ref{bbk1}.
\end{theorem}

\begin{IEEEproof}
Since $\mathcal{C}_{s,gmac-dms}^{f}\subseteq\mathcal{C}_{gmac-dms}^{f}=\mathcal{C}_{gmac-dms}$,
we only need to show that any achievable rate pair $(R_{1},R_{2})$ in
$\mathcal{C}_{gmac-dms}$ satisfies the secrecy constraint in (\ref{wawa2}).

In the preceding subsection, we introduce a two-step SK scheme for the GMAC-DMS with feedback, and show that this scheme achieves $\mathcal{C}_{gmac-dms}^{f}$.
In this new scheme, the transmitted codewords $X_{1,i}$, $U_{i}$ and $V_{i}$ at time $i$ ($1\leq i\leq N$) can be expressed as
\begin{eqnarray}\label{gxt1}
&&X_{1,1}=0,\,\,U_{1}=0,\,\,V_{1}=\sqrt{12(1-\rho^{2})P_{2}}\theta_{2},\nonumber\\
&&X_{1,2}=\frac{\sqrt{12P^{*}}\theta_{1}}{\rho\sqrt{\frac{P_{2}}{P_{1}}}+1},\,\,U_{2}=\rho\sqrt{\frac{P_{2}}{P_{1}}}X_{1,2},\,\,
V_{2}=\sqrt{\frac{(1-\rho^{2})P_{2}}{\sigma_{1}^{2}}}\eta_{1,1},\nonumber\\
&&X_{1,3}=\frac{\sqrt{P^{*}P_{2}(1-\rho^{2})}}{\sigma_{1}r(\rho\sqrt{\frac{P_{2}}{P_{1}}}+1)}\eta_{1,1}+
\frac{\sqrt{P^{*}}}{r(\rho\sqrt{\frac{P_{2}}{P_{1}}}+1)}\eta_{1,2},\,\,U_{3}=\rho\sqrt{\frac{P_{2}}{P_{1}}}X_{1,3},\nonumber\\
&&V_{3}=\frac{\sqrt{(1-\rho^{2})P_{2}}}{r}\eta_{1,1}-\frac{(1-\rho^{2})P_{2}}{r\sigma_{1}}\eta_{1,2},\nonumber\\
&&...\nonumber\\
&&X_{1,N}=\frac{1}{\rho\sqrt{\frac{P_{2}}{P_{1}}}+1}
\sqrt{\frac{P^{*}}{\alpha^{'}_{N-1}}}\left(\epsilon^{'}_{N-2}\frac{r^{2}}{P^{*}+r^{2}}-(\eta_{1,N-1}+
\sqrt{\frac{\alpha_{N-3}}{\alpha_{N-2}}}\frac{\sigma_{1}^{2}}{r^{2}}V_{N-2}\right.\nonumber\\
&&\left.-\sqrt{\frac{\alpha_{N-3}}{\alpha_{N-2}}}\frac{(1-\rho^{2})P_{2}}{r^{2}}\eta_{1,N-2})\cdot
\frac{\sqrt{P^{*}\cdot\alpha^{'}_{N-2}}}{P^{*}+r^{2}}\right)
,\,\,U_{N}=\rho\sqrt{\frac{P_{2}}{P_{1}}}X_{1,N}\nonumber\\
&&V_{N}=\sqrt{\frac{\alpha_{N-2}}{\alpha_{N-1}}}\frac{\sigma_{1}^{2}}{r^{2}}V_{N-1}
-\sqrt{\frac{\alpha_{N-2}}{\alpha_{N-1}}}\frac{(1-\rho^{2})P_{2}}{r^{2}}\eta_{1,N-1},
\end{eqnarray}
where $r$ is defined in (\ref{qqe1.xxs24}) and $P^{*}$ is defined in (\ref{clss4.rr3}).

From (\ref{gxt1}), we can conclude that for $3\leq k\leq N$,
$X_{1,k}$, $U_{k}$ and $V_{k}$ are functions of $\eta_{1,1}$,...,$\eta_{1,k-1}$, and they are independent of the transmitted messages.
For convenience, for $3\leq k\leq N$, define
\begin{eqnarray}\label{quanyou1}
&&X_{1,k}=f_{1,k}(\eta_{1,1},...,\eta_{1,k-1}),\,\,\,U_{k}=f_{u,k}(\eta_{1,1},...,\eta_{1,k-1}),\,\,\,V_{k}=f_{v,k}(\eta_{1,1},...,\eta_{1,k-1}).
\end{eqnarray}
By using (\ref{gxt1}) and (\ref{quanyou1}), the equivocation rate
$\frac{1}{N}H(W_{1},W_{2}|Z^{N})$ of any achievable rate pair in $\mathcal{C}_{gmac-dms}^{f}$ can be bounded by
\begin{eqnarray}\label{quanyou2}
&&\frac{1}{N}H(W_{1},W_{2}|Z^{N})=\frac{1}{N}H(\theta_{1},\theta_{2}|Z^{N})\nonumber\\
&&\geq\frac{1}{N}H(\theta_{1},\theta_{2}|Z^{N},\eta_{1,1},...,\eta_{1,N},\eta_{2,3},...,\eta_{2,N})\nonumber\\
&&\stackrel{(a)}=\frac{1}{N}H(\theta_{1},\theta_{2}|\sqrt{12(1-\rho^{2})P_{2}}\theta_{2}+\eta_{1,1}+\eta_{2,1},
\sqrt{12P^{*}}\theta_{1}+\sqrt{\frac{(1-\rho^{2})P_{2}}{\sigma_{1}^{2}}}\eta_{1,1}+\eta_{1,2}+\eta_{2,2},\nonumber\\
&&f_{1,3}(\eta_{1,1},\eta_{1,2})+f_{u,3}(\eta_{1,1},\eta_{1,2})+f_{v,3}(\eta_{1,1},\eta_{1,2})+\eta_{1,3}+\eta_{2,3},...,f_{1,N}(\eta_{1,1},...,\eta_{1,N-1})\nonumber\\
&&+f_{u,N}(\eta_{1,1},...,\eta_{1,N-1})+f_{v,N}(\eta_{1,1},...,\eta_{1,N-1})
+\eta_{1,N}+\eta_{2,N},\eta_{1,1},...,\eta_{1,N},\eta_{2,3},...,\eta_{2,N})\nonumber\\
&&\stackrel{(b)}=\frac{1}{N}H(\theta_{1},\theta_{2}|\sqrt{12(1-\rho^{2})P_{2}}\theta_{2}+\eta_{2,1},
\sqrt{12P^{*}}\theta_{1}+\eta_{2,2})\nonumber\\
&&=\frac{1}{N}(H(\theta_{1},\theta_{2})-h(\sqrt{12(1-\rho^{2})P_{2}}\theta_{2}+\eta_{2,1},
\sqrt{12P^{*}}\theta_{1}+\eta_{2,2})
+h(\eta_{2,1},\eta_{2,2}|\theta_{1},\theta_{2}))\nonumber\\
&&\stackrel{(c)}=\frac{1}{N}(H(\theta_{1},\theta_{2})-h(\sqrt{12(1-\rho^{2})P_{2}}\theta_{2}+\eta_{2,1})
-h(\sqrt{12P^{*}}\theta_{1}+\eta_{2,2})+h(\eta_{2,1})+h(\eta_{2,2}))\nonumber\\
&&\stackrel{(d)}\geq R_{1}+R_{2}-(\frac{1}{2N}\log(1+\frac{(1-\rho^{2})P_{2}}{\sigma_{2}^{2}})+\frac{1}{2N}\log(1+\frac{P^{*}}{\sigma_{2}^{2}})),
\end{eqnarray}
where (a) follows from (\ref{wawa1}), (\ref{gxt1}) and (\ref{quanyou1}), (b) follows from the fact that $\theta_{1}$, $\theta_{2}$, $\eta_{2,1}$ and $\eta_{2,2}$ are independent of
$\eta_{1,1},...,\eta_{1,N}$, $\eta_{2,3},...,\eta_{2,N}$,
(c) follows from the fact that $\theta_{1}$, $\theta_{2}$, $\eta_{2,1}$ and $\eta_{2,2}$ are independent of
each other, and (d) follows because $H(\theta_{j})=NR_{j}$ ($j=1,2$), the variance of $\theta_{j}$ equals $\frac{1}{12}$
as $N$ tends to infinity, and $\theta_{j}$ is independent of $\eta_{2,j}$.
Choosing sufficiently large $N$, the secrecy constraint in (\ref{wawa2}) is guaranteed, which indicates that
any achievable rate pair $(R_{1},R_{2})$ in
$\mathcal{C}_{gmac-dms}^{f}$ is achievable with perfect weak secrecy, and hence $\mathcal{C}_{s,gmac-dms}^{f}=\mathcal{C}_{gmac-dms}^{f}$.
The proof of Theorem \ref{T5} is completed.

\end{IEEEproof}

For comparison,
the following Corollary \ref{T6} establishes
an outer bound on the secrecy capacity region $\mathcal{C}_{s,gmac-dms}$ of GMAC-WT-DMS \emph{without feedback}.

\begin{corollary}\label{T6}
$\mathcal{C}_{s,gmac-dms}\subseteq\mathcal{C}^{out}_{s,gmac-dms}$, where $\mathcal{C}^{out}_{s,gmac-dms}$ is given by
\begin{eqnarray}\label{liqin1.r}
&&\mathcal{C}^{out}_{s,gmac-dms}=\bigcup_{-1\leq\rho\leq 1}\{(R_{1}\geq 0,R_{2}\geq 0):
R_{2}\leq\frac{1}{2}\log\left(1+\frac{(1-\rho^{2})P_{2}}{\sigma_{1}^{2}}\right),\nonumber\\
&&R_{1}+R_{2}\leq \frac{1}{2}\log\left(1+\frac{P_{1}+P_{2}+2\sqrt{P_{1}P_{2}}\rho}{\sigma_{1}^{2}}\right)
-\frac{1}{2}\log\left(1+\frac{P_{1}+P_{2}+2\sqrt{P_{1}P_{2}}\rho}{\sigma_{1}^{2}+\sigma_{2}^{2}}\right)\}.\nonumber\\
\end{eqnarray}
\end{corollary}
\begin{IEEEproof}
See Appendix \ref{appen1}.
\end{IEEEproof}

The following Figure \ref{f13} shows the rate gains by using channel feedback for
$P_{1}=1$, $P_{2}=1.5$, $\sigma^{2}_{1}=0.1$ and $\sigma_{2}^{2}=1.2$.

\begin{figure}[htb]
\centering
\includegraphics[scale=0.5]{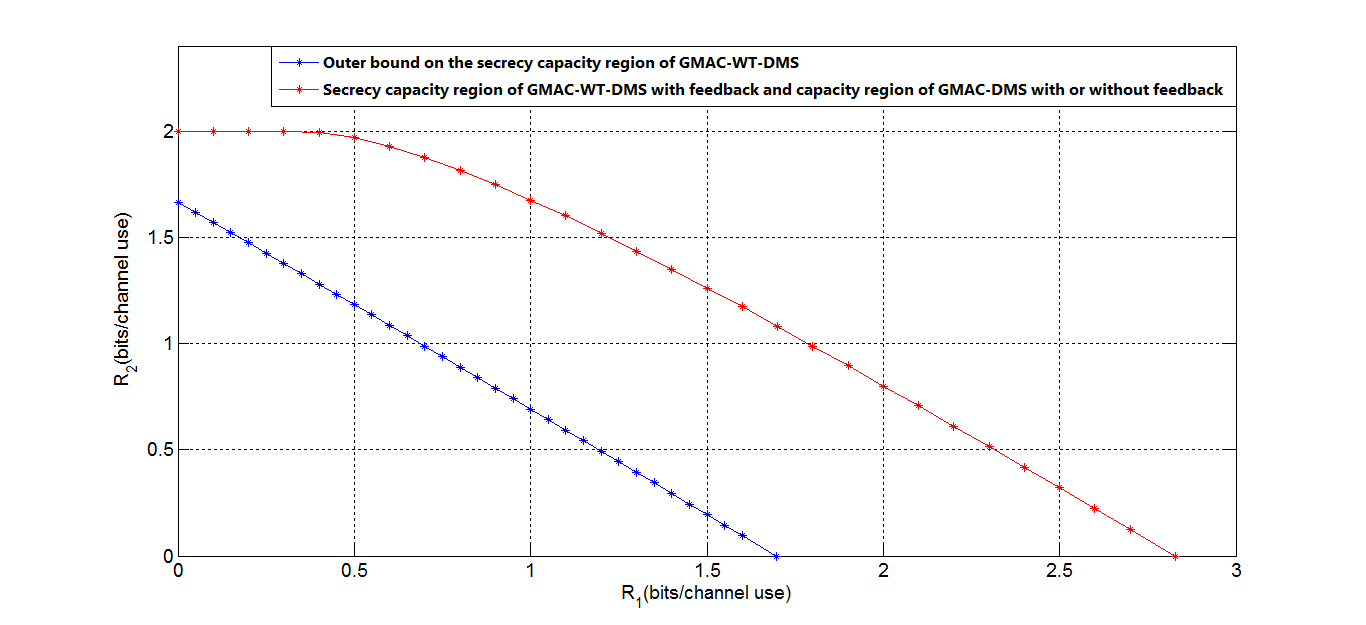}
\caption{Capacity results on GMAC-WT-DMS with or without feedback}
\label{f13}
\end{figure}

\section{The SSCA feedback scheme for the GMAC-WT-NCSIT-DMS}\label{secIV}
\setcounter{equation}{0}

In this section, first, we extend the two-step SK-type
feedback scheme of the preceding section to the GMAC-NCSIT-DMS with feedback, and show that this extended scheme is also capacity-achieving.
Second, we show that this extended feedback scheme is secure by itself and also achieves
the secrecy capacity region $\mathcal{C}^{f}_{s,gmac-ncsit-dms}$ of the GMAC-WT-NCSIT-DMS with feedback.
Finally, in order to show the
rate gains by the feedback, an outer bound on the secrecy capacity region $\mathcal{C}_{s,gmac-ncsit-dms}$ of the GMAC-WT-NCSIT-DMS is provided,
and the capacity results given in this section are further explained via a numerical example.

\subsection{A capacity-achieving SK-type scheme for the GMAC-NCSIT-DMS with feedback}\label{secIV-1}

The model of the GMAC-NCSIT-DMS with feedback is formulated in Section \ref{secII-y2-yy1}.
In this subsection, first, we introduce capacity results on the GMAC-NCSIT-DMS with or without feedback. Then, we propose a corresponding 
capacity-achieving feedback scheme.

\subsubsection{Capacity results on the GMAC-NCSIT-DMS with or without feedback}\label{secIV-1-1}

The following Corollary \ref{bbk2} characterizes the capacity region $\mathcal{C}_{gmac-ncsit-dms}$ of the GMAC-NCSIT-DMS.

\begin{corollary}\label{bbk2}
The capacity region $\mathcal{C}_{gmac-ncsit-dms}$ of the GMAC-NCSIT-DMS is given by
\begin{eqnarray}\label{x.clss1}
&&\mathcal{C}_{gmac-ncsit-dms}=\mathcal{C}_{gmac-dms}=\bigcup_{0\leq \rho\leq 1}\left\{(R_{1},R_{2}):R_{2}\leq\frac{1}{2}\log\left(1+\frac{P_{2}(1-\rho^{2})}{\sigma_{1}^{2}}\right),\right.\nonumber\\
&&\left.R_{1}+R_{2}\leq \frac{1}{2}\log\left(1+\frac{P_{1}+P_{2}+2\sqrt{P_{1}P_{2}}\rho}{\sigma_{1}^{2}}\right)\right\}.
\end{eqnarray}
\end{corollary}

\begin{IEEEproof}
In \cite{kim1}, it has been pointed out that $\mathcal{C}_{gmac-ncsit-dms}$
equals $\mathcal{C}_{gmac-dms}$, which indicates that for the GMAC-NCSIT-DMS, the state interference can be pre-cancelled by both the transmitters and the receiver.

\end{IEEEproof}

The following Corollary \ref{T7} determines the capacity region $\mathcal{C}^{f}_{gmac-ncsit-dms}$ of the GMAC-NCSIT-DMS with feedback, which indicates that
feedback does not increase the capacity of the GMAC-NCSIT-DMS, see the followings.

\begin{corollary}\label{T7}
$\mathcal{C}^{f}_{gmac-ncsit-dms}=\mathcal{C}_{gmac-ncsit-dms}$, where $\mathcal{C}_{gmac-ncsit-dms}$ is given in (\ref{x.clss1}).
\end{corollary}

\begin{IEEEproof}
Note that $\mathcal{C}^{f}_{gmac-ncsit-dms}\subseteq\mathcal{C}^{f}_{gmac-dms}=\mathcal{C}_{gmac-dms}=\mathcal{C}_{gmac-ncsit-dms}$
directly follows from the converse proof of the bounds on $R_{2}$ and $R_{1}+R_{2}$
in $\mathcal{C}^{f}_{gmac}$ \cite[pp. 627-628]{cover7}, and hence we omit the converse proof here.
On the other hand, note that $\mathcal{C}_{gmac-ncsit-dms}\subseteq \mathcal{C}^{f}_{gmac-ncsit-dms}$
since non-feedback model is a special case of the feedback model,
and $\mathcal{C}_{gmac-ncsit-dms}$\\$=\mathcal{C}_{gmac-dms}$ (see (\ref{x.clss1})),
and hence the proof of Theorem \ref{T7} is completed.

\end{IEEEproof}

\subsubsection{A capacity-achieving two-step SK-type feedback scheme for the GMAC-NCSIT-DMS with feedback}\label{secIV-1-2}

Similar to the two-step SK-type feedback scheme in Section \ref{secIII},
Transmitter $1$ uses power $P_{1}$ to encode $W_{1}$, $S^{N}$ and the feedback $Y^{N}$ as $X_{1}^{N}$. Transmitter $2$ uses power $(1-\rho^{2})P_{2}$
to encode $W_{2}$, $S^{N}$ and $Y^{N}$ as $V^{N}$, and power $\rho^{2}P_{2}$ to encode $W_{1}$, $S^{N}$ and $Y^{N}$ as $U^{N}$, where $0\leq \rho \leq 1$.
Moreover, let $\theta_{j}$ be the center of the sub-interval w.r.t. the message $W_{j}$
(the variance of $\theta_{j}$ approximately equals $\frac{1}{12}$).

\emph{Encoding:}
At time $1$, Transmitter $1$ sends
\begin{eqnarray}\label{x.sheva1}
&&X_{1,1}=0.
\end{eqnarray}
Transmitter $2$ sends
\begin{eqnarray}\label{r.qqe1}
&&V_{1}=\sqrt{12(1-\rho^{2})P_{2}}(\theta_{2}-\frac{S_{1}}{\sqrt{12(1-\rho^{2})P_{2}}}+A_{2}),
\end{eqnarray}
and
\begin{eqnarray}\label{x.sheva2}
&&U_{1}=\rho\sqrt{\frac{P_{2}}{P_{1}}}X_{1,1}=0,
\end{eqnarray}
where $A_{2}$ is a linear combination of $S_{1}$,...,$S_{N}$, and it will be determined later.

The receiver obtains
\begin{eqnarray}\label{dankerxx.sheva2}
&&Y_{1}=V_{1}+X_{1,1}+U_{1}+S_{1}+\eta_{1,1}=V_{1}+S_{1}+\eta_{1,1}=\sqrt{12(1-\rho^{2})P_{2}}\theta_{2}+\sqrt{12(1-\rho^{2})P_{2}}A_{2}+\eta_{1,1},\nonumber\\
\end{eqnarray}
and gets an estimation $\hat{\theta}_{2,1}$ of $\theta_{2}$ by computing
\begin{eqnarray}\label{dxx.sheva2}
&&\hat{\theta}_{2,1}=\frac{Y_{1}}{\sqrt{12(1-\rho^{2})P_{2}}}=\theta_{2}+A_{2}+\frac{\eta_{1,1}}{\sqrt{12(1-\rho^{2})P_{2}}}=\theta_{2}+A_{2}+\epsilon_{1},
\end{eqnarray}
where $\epsilon_{1}$ is in the same fashion as that in Section \ref{secIII}, and define
$\alpha_{1}\triangleq Var(\epsilon_{1})=\frac{\sigma_{1}^{2}}{12(1-\rho^{2})P_{2}}$. Then the receiver
sends $Y_{1}$ back to Transmitter $2$. Let $Y^{'}_{1}=Y_{1}=V_{1}+S_{1}+\eta_{1,1}$,
Transmitter $2$ computes
\begin{eqnarray}\label{dx.qqe2}
&&\frac{Y^{'}_{1}}{\sqrt{12(1-\rho^{2})P_{2}}}=\theta_{2}+A_{2}+\frac{\eta_{1,1}}{\sqrt{12(1-\rho^{2})P_{2}}}=\theta_{2}+A_{2}+\epsilon_{1}.
\end{eqnarray}
Since $A_{2}$ is known by the transmitters, Transmitter $2$ obtains $\epsilon_{1}$ from (\ref{dx.qqe2}).

At time $2$, Transmitter $2$ sends $V_{2}$ exactly in the same fashion as that in (\ref{qqe1.xxs1}), i.e.,
$V_{2}=\sqrt{\frac{(1-\rho^{2})P_{2}}{\alpha_{1}}}\epsilon_{1}$.
On the other hand, at time $2$,
Transmitters $1$ and $2$ respectively send $X_{1,2}$ and $U_{2}=\rho\sqrt{\frac{P_{2}}{P_{1}}}X_{1,2}$ such that
\begin{eqnarray}\label{xxs.qqe1.r1}
&&X_{2}^{*}=U_{2}+X_{1,2}=\sqrt{12P^{*}}(\theta_{1}-\frac{S_{2}}{\sqrt{12P^{*}}}+A_{1}),
\end{eqnarray}
where $P^{*}$ is defined in the same fashion as that in (\ref{clss4.rr3}) and
\begin{eqnarray}\label{xxs.xx.sheva2}
&&A_{1}=\sum_{i=3}^{N}\beta_{1,i}S_{i},
\end{eqnarray}
and $\beta_{1,i}$ will be defined later.
The receiver obtains
\begin{eqnarray}\label{gg.dankerxx.sheva2}
&&Y_{2}=X_{2}^{*}+V_{2}+S_{2}+\eta_{1,2}=\sqrt{12P^{*}}\theta_{1}+\sqrt{12P^{*}}A_{1}+V_{2}+\eta_{1,2},
\end{eqnarray}
and gets an estimation $\hat{\theta}_{1,2}$ of $\theta_{1}$ by computing
\begin{eqnarray}\label{gg.dxx.sheva2}
&&\hat{\theta}_{1,2}=\frac{Y_{2}}{\sqrt{12P^{*}}}=\theta_{1}+A_{1}+\frac{V_{2}+\eta_{1,2}}{\sqrt{12P^{*}}}=\theta_{1}+A_{1}+\epsilon^{'}_{2},
\end{eqnarray}
where $\epsilon^{'}_{2}$ is in the same fashion as that in Section \ref{secIII}, and define
$\alpha^{'}_{2}\triangleq Var(\epsilon^{'}_{2})$. Then the receiver
sends $Y_{2}$ back to both transmitters.

At time $3$, once receiving the feedback $Y_{2}=X_{2}^{*}+V_{2}+S_{2}+\eta_{1,2}$, both transmitters compute
\begin{eqnarray}\label{xxs.qqe2.r2}
&&\frac{Y_{2}}{\sqrt{12P^{*}}}=\theta_{1}+A_{1}+\frac{V_{2}+\eta_{1,2}}{\sqrt{12P^{*}}}=\theta_{1}+A_{1}+\epsilon^{'}_{2}.
\end{eqnarray}
and send $X_{1,3}$ and $U_{3}=\rho\sqrt{\frac{P_{2}}{P_{1}}}X_{1,3}$ such that
\begin{eqnarray}\label{xxs.sheva5}
&&X_{3}^{*}=U_{3}+X_{1,3}=\sqrt{\frac{P^{*}}{\alpha^{'}_{2}}}\epsilon^{'}_{2}.
\end{eqnarray}
In addition, subtracting $X_{1,2}$, $U_{2}$ and $S_{2}$ from $Y_{2}$ and let
$Y^{'}_{2}=Y_{2}-X_{1,2}-U_{2}-S_{2}=V_{2}+\eta_{1,2}$,
Transmitter $2$ computes
\begin{eqnarray}\label{xxs.sheva3}
&&\epsilon_{2}=\epsilon_{1}-\frac{E[Y^{'}_{2}\epsilon_{1}]}{E[(Y^{'}_{2})^{2}]}Y^{'}_{2}.
\end{eqnarray}
and sends
\begin{eqnarray}\label{xxs.sheva4}
&&V_{3}=\sqrt{\frac{(1-\rho^{2})P_{2}}{\alpha_{2}}}\epsilon_{2},
\end{eqnarray}
where $\alpha_{2}\triangleq Var(\epsilon_{2})$.

At time $4\leq k\leq N$, once receiving $Y_{k-1}=X_{1,k-1}+U_{k-1}+V_{k-1}+S_{k}+\eta_{1,k-1}$,
Transmitter $2$ computes
\begin{eqnarray}\label{xxs.qqe1.xxs3}
&&\epsilon_{k-1}=\epsilon_{k-2}-\beta_{2,k-1}Y^{'}_{k-1},
\end{eqnarray}
where
\begin{eqnarray}\label{xxs.sheva6}
&&Y^{'}_{k-1}=Y_{k-1}-X_{1,k-1}-U_{k-1}-S_{k-1},
\end{eqnarray}
\begin{eqnarray}\label{umg.xxs.sheva6}
&&\beta_{2,k-1}=\frac{E[Y^{'}_{k-1}\epsilon_{k-2}]}{E[(Y^{'}_{k-1})^{2}]},
\end{eqnarray}
and sends
\begin{eqnarray}\label{r.qqe1.xxs4}
&&V_{k}=\sqrt{\frac{(1-\rho^{2})P_{2}}{\alpha_{k-1}}}\epsilon_{k-1},
\end{eqnarray}
where $\alpha_{k-1}\triangleq Var(\epsilon_{k-1})$.
In the meanwhile, Transmitters $1$ and $2$ respectively send $X_{1,k}$ and $U_{k}=\rho\sqrt{\frac{P_{2}}{P_{1}}}X_{1,k}$ such that
\begin{eqnarray}\label{xxs.qqe1.xxs5}
&&X_{k}^{*}=U_{k}+X_{1,k}=\sqrt{\frac{P^{*}}{\alpha^{'}_{k-1}}}\epsilon^{'}_{k-1},
\end{eqnarray}
where
\begin{eqnarray}\label{xxs.qqe1.xxs6}
&&\epsilon^{'}_{k-1}=\epsilon^{'}_{k-2}-\beta_{1,k-1}(Y_{k-1}-S_{k-1}),
\end{eqnarray}
\begin{eqnarray}\label{umg.xxs.qqe1.xxs6}
&&\beta_{1,k-1}=\frac{E[(Y_{k-1}-S_{k-1})\epsilon^{'}_{k-2}]}{E[(Y_{k-1}-S_{k-1})^{2}]},
\end{eqnarray}
and $\alpha^{'}_{k-1}\triangleq Var(\epsilon^{'}_{k-1})$.

Here note that though
the use of $S^{N}$ at time instants $1$ and $2$ causes the transmission power of the first two time instants
to be larger than the average power constraint. However, for $k\geq 3$, the transmission power equals the average power constraint (see the analysis of the
average transmission power in Section \ref{secIII-1-2}), and hence for
sufficiently larger $N$, the power constraint is preserved.

\emph{Decoding:}

The receiver uses a two-step decoding scheme which is similar to that in Section \ref{secIII}. Specifically,
first, from (\ref{qe8}), we observe that at time $k$ ($3\leq k\leq N$),
the receiver's estimation $\hat{\theta}_{1,k}$ of $\theta_{1}$ is given by
\begin{eqnarray}\label{xxs.qqe1.xxs7}
&&\hat{\theta}_{1,k}=\hat{\theta}_{1,k-1}-\beta_{1,k}Y_{k},
\end{eqnarray}
where $\beta_{1,k}=\frac{E[(Y_{k}-S_{k})\epsilon^{'}_{k-1}]}{E[(Y_{k}-S_{k})^{2}]}$.
Combining (\ref{xxs.qqe1.xxs6}) with (\ref{xxs.qqe1.xxs7}), we have
\begin{eqnarray}\label{xxs.qqe1.xxs7.xx1}
\hat{\theta}_{1,k}&=&\hat{\theta}_{1,k-1}+\epsilon^{'}_{k}-\epsilon^{'}_{k-1}-\beta_{1,k}S_{k}\nonumber\\
&=&\hat{\theta}_{1,2}+\epsilon^{'}_{k}-\epsilon^{'}_{2}-\sum_{j=3}^{k}\beta_{1,j}S_{j}\nonumber\\
&\stackrel{(a)}=&\theta_{1}+A_{1}+\epsilon^{'}_{2}+\epsilon^{'}_{k}-\epsilon^{'}_{2}-\sum_{j=3}^{k}\beta_{1,j}S_{j}\nonumber\\
&=&\theta_{1}+\epsilon^{'}_{k}+A_{1}-\sum_{j=3}^{k}\beta_{1,j}S_{j},
\end{eqnarray}
where (a) follows from (\ref{gg.dxx.sheva2}). From (\ref{xxs.qqe1.xxs7.xx1}), we can conclude that for $k=N$,
\begin{eqnarray}\label{xxs.qqe1.xxs7.xx1.dlrb1}
\hat{\theta}_{1,N}&=&\theta_{1}+\epsilon^{'}_{N}+A_{1}-\sum_{j=3}^{N}\beta_{1,j}S_{j}\nonumber\\
&\stackrel{(b)}=&\theta_{1}+\epsilon^{'}_{N}+A_{1}-A_{1}=\theta_{1}+\epsilon^{'}_{N},
\end{eqnarray}
where (b) follows from (\ref{xxs.xx.sheva2}).
Note that (\ref{xxs.qqe1.xxs7.xx1.dlrb1})
indicates that the receiver's final estimation of $\theta_{1}$ is in the same fashion as that in Section \ref{secIII}, and
observing that $\epsilon^{'}_{k}$ ($2\leq k\leq N$) is exactly in the same fashion as those in Section \ref{secIII}, we can directly apply
Lemma \ref{L1} to show that the decoding error probability $P_{e1}$ of $\theta_{1}$ tends to $0$ as $N\rightarrow \infty$ if
$R_{1}<\frac{1}{2}\log(1+\frac{P_{1}+\rho^{2}P_{2}+2\sqrt{P_{1}P_{2}}\rho}{(1-\rho^{2})P_{2}+\sigma_{1}^{2}})$ is satisfied.

Second, after decoding $W_{1}$ ($\theta_{1}$), the receiver obtains $\epsilon^{'}_{k}+A_{1}-\sum_{j=3}^{k}\beta_{1,j}S_{j}$ ($3\leq k\leq N$)
from (\ref{xxs.qqe1.xxs7.xx1}), and obtains $\epsilon^{'}_{2}+A_{1}$ from (\ref{gg.dxx.sheva2}).
Furthermore, from (\ref{xxs.qqe1.xxs5}) and the fact that $\sqrt{\frac{P^{*}}{\alpha^{'}_{k}}}$ is a constant value, we can conclude that for $3\leq k\leq N$,
the receiver knows
\begin{eqnarray}\label{bg1}
&&\sqrt{\frac{P^{*}}{\alpha^{'}_{k}}}(\epsilon^{'}_{k}+A_{1}-\sum_{j=3}^{k}\beta_{1,j}S_{j})=
X^{*}_{k+1}+\sqrt{\frac{P^{*}}{\alpha^{'}_{k}}}(A_{1}-\sum_{j=3}^{k}\beta_{1,j}S_{j}).
\end{eqnarray}
In addition,
for $k=2$, the receiver knows
\begin{eqnarray}\label{bg2}
&&X^{*}_{3}+\sqrt{\frac{P^{*}}{\alpha^{'}_{2}}}A_{1}
\end{eqnarray}
since $X^{*}_{3}=\sqrt{\frac{P^{*}}{\alpha^{'}_{2}}}\epsilon^{'}_{2}$ and $\sqrt{\frac{P^{*}}{\alpha^{'}_{2}}}$ is a constant value.
Here for $k=2$, define $\sum_{j=3}^{k}\beta_{1,j}S_{j}=0$. Then we can conclude that the receiver knows the terms in (\ref{bg1})
for $2\leq k\leq N$. Here note that in the decoding procedure of the two-step SK-type scheme for the GMAC-DMS, after decoding $\theta_{1}$, the receiver knows
$X_{1,k}$ and $U_{k}$ for all $3\leq k\leq N$, and hence he/she subtracts $X_{1,k}$ and $U_{k}$ from $Y_{k}$, and further using SK-type decoding
scheme to obtain $\theta_{2}$. While, in this extended scheme for the GMAC-NCSIT-DMS, after decoding $\theta_{1}$, the receiver does not know
$X_{1,k}$ and $U_{k}$, instead, he/she only knows $X^{*}_{k}+\sqrt{\frac{P^{*}}{\alpha^{'}_{k-1}}}(A_{1}-\sum_{j=3}^{k-1}\beta_{1,j}S_{j})$,
then the key to further decode $\theta_{2}$ is how to choose $A_{2}$ (a linear combination of $(S_{1},...,S_{N})$) to precancel
the offset of the receiver's final estimation of $\theta_{2}$.

Recall that the receiver's estimation $\hat{\theta}_{2,1}$ of $\theta_{2}$ is given by (\ref{dxx.sheva2}). At time $2$, since
$\theta_{1}$ is obtained by the receiver,
the receiver's estimation $\hat{\theta}_{2,2}$ of $\theta_{2}$ is given by
\begin{eqnarray}\label{bg3}
\hat{\theta}_{2,2}&=&\hat{\theta}_{2,1}-\beta_{2,2}(Y_{2}-\sqrt{12P^{*}}\theta_{1})\nonumber\\
&\stackrel{(c)}=&\hat{\theta}_{2,1}+\epsilon_{2}-\epsilon_{1}-\beta_{2,2}\sqrt{12P^{*}}A_{1}\nonumber\\
&\stackrel{(d)}=&\theta_{2}+A_{2}+\epsilon_{1}+\epsilon_{2}-\epsilon_{1}-\beta_{2,2}\sqrt{12P^{*}}A_{1}=\theta_{2}+\epsilon_{2}+A_{2}-\beta_{2,2}\sqrt{12P^{*}}A_{1},
\end{eqnarray}
where (c) follows from (\ref{xxs.sheva3}), and (d) follows from (\ref{dxx.sheva2}). At time $k$ ($3\leq k\leq N$),
the receiver's estimation $\hat{\theta}_{2,k}$ of $\theta_{2}$ is given by
\begin{eqnarray}\label{bg4}
\hat{\theta}_{2,k}&\stackrel{(e)}=&\hat{\theta}_{2,k-1}-\beta_{2,k}\left(Y_{k}-X^{*}_{k}-\sqrt{\frac{P^{*}}{\alpha^{'}_{k-1}}}(A_{1}
-\sum_{j=3}^{k-1}\beta_{1,j}S_{j})\right)\nonumber\\
&\stackrel{(f)}=&\hat{\theta}_{2,k-1}+\epsilon_{k}-\epsilon_{k-1}-\beta_{2,k}S_{k}+\beta_{2,k}\sqrt{\frac{P^{*}}{\alpha^{'}_{k-1}}}\left(A_{1}
-\sum_{j=3}^{k-1}\beta_{1,j}S_{j}\right)\nonumber\\
&=&\hat{\theta}_{2,2}+\epsilon_{k}-\epsilon_{2}+\sum_{i=3}^{k}\left(\beta_{2,i}\sqrt{\frac{P^{*}}{\alpha^{'}_{i-1}}}(A_{1}
-\sum_{j=3}^{i-1}\beta_{1,j}S_{j})-\beta_{2,i}S_{i}\right)\nonumber\\
&\stackrel{(g)}=&\theta_{2}+\epsilon_{2}+A_{2}-\beta_{2,2}\sqrt{12P^{*}}A_{1}+\epsilon_{k}-\epsilon_{2}+\sum_{i=3}^{k}\left(\beta_{2,i}\sqrt{\frac{P^{*}}{\alpha^{'}_{i-1}}}(A_{1}
-\sum_{j=3}^{i-1}\beta_{1,j}S_{j})-\beta_{2,i}S_{i}\right)\nonumber\\
&=&\theta_{2}+\epsilon_{k}+A_{2}-\beta_{2,2}\sqrt{12P^{*}}A_{1}+\sum_{i=3}^{k}\left(\beta_{2,i}\sqrt{\frac{P^{*}}{\alpha^{'}_{i-1}}}(A_{1}
-\sum_{j=3}^{i-1}\beta_{1,j}S_{j})-\beta_{2,i}S_{i}\right),
\end{eqnarray}
where (e) follows from the fact that the term in (\ref{bg1}) is known by the receiver and hence it can be subtracted from $Y_{k}$, (f) follows from (\ref{xxs.qqe1.xxs3}), and
(g) follows from (\ref{bg3}). From (\ref{bg4}), we can conclude that for $k=N$,
\begin{eqnarray}\label{bg5}
&&\hat{\theta}_{2,N}=\theta_{2}+\epsilon_{N}+A_{2}-\beta_{2,2}\sqrt{12P^{*}}A_{1}+\sum_{i=3}^{N}\left(\beta_{2,i}\sqrt{\frac{P^{*}}{\alpha^{'}_{i-1}}}(A_{1}
-\sum_{j=3}^{i-1}\beta_{1,j}S_{j})-\beta_{2,i}S_{i}\right).\nonumber\\
\end{eqnarray}
Observing that if
\begin{eqnarray}\label{bg6}
A_{2}=\beta_{2,2}\sqrt{12P^{*}}A_{1}-\sum_{i=3}^{N}\left(\beta_{2,i}\sqrt{\frac{P^{*}}{\alpha^{'}_{i-1}}}(A_{1}
-\sum_{j=3}^{i-1}\beta_{1,j}S_{j})-\beta_{2,i}S_{i}\right),
\end{eqnarray}
(\ref{bg5}) can be re-written as
\begin{eqnarray}\label{bg7}
&&\hat{\theta}_{2,N}=\theta_{2}+\epsilon_{N},
\end{eqnarray}
which indicates that the receiver's final estimation of $\theta_{2}$ is in the same fashion as that in Section \ref{secIII}, and
observing that $\epsilon_{k}$ ($1\leq k\leq N$) is exactly in the same fashion as those in Section \ref{secIII}, we can directly apply the same
argument in Section \ref{secIII} to show that
the decoding error probability $P_{e2}$ of $\theta_{2}$ tends to $0$ as $N\rightarrow \infty$ if
$R_{2}<\frac{1}{2}\log(1+\frac{(1-\rho^{2})P_{2}}{\sigma_{1}^{2}})$ is satisfied.

Finally, note that the decoding error probability $P_{e}$ of the receiver is upper bounded by
$P_{e}\leq P_{e1}+P_{e2}$, and from above analysis, we can conclude that
the rate pair $(R_{1}=\frac{1}{2}\log(1+\frac{P_{1}+\rho^{2}P_{2}+2\sqrt{P_{1}P_{2}}\rho}{(1-\rho^{2})P_{2}+\sigma_{1}^{2}}),
R_{2}=\frac{1}{2}\log(1+\frac{(1-\rho^{2})P_{2}}{\sigma_{1}^{2}}))$ is achievable for all $0\leq \rho\leq 1$, which indicates that all rate pairs $(R_{1},R_{2})$
in $\mathcal{C}^{f}_{gmac-ncsit-dms}$ are achievable. Hence this extended two-step SK-type feedback scheme achieves the capacity region
$\mathcal{C}^{f}_{gmac-ncsit-dms}$ of GMAC-NCSIT-DMS with feedback.

\subsection{Capacity results on the GMAC-WT-NCSIT-DMS with or without feedback}\label{secIV-2}

The model of the GMAC-WT-NCSIT-DMS with feedback is formulated in Section \ref{secII-y2-yy3}. The following Theorem \ref{T8}
establishes that the secrecy constraint does not reduce the capacity of GMAC-NCSIT-DMS with feedback.

\begin{theorem}\label{T8}
$\mathcal{C}_{s,gmac-ncsit-dms}^{f}=\mathcal{C}_{gmac-ncsit-dms}^{f}$, where $\mathcal{C}_{s,gmac-ncsit-dms}^{f}$ is the secrecy capacity region of the
GMAC-WT-NCSIT-DMS with feedback, and
$\mathcal{C}_{gmac-ncsit-dms}^{f}$ is given in Corollary \ref{T7}.
\end{theorem}

\begin{IEEEproof}
Since $\mathcal{C}_{s,gmac-ncsit-dms}^{f}\subseteq\mathcal{C}_{gmac-ncsit-dms}^{f}$, we only need to show that any achievable rate pair $(R_{1},R_{2})$ in
$\mathcal{C}_{gmac-ncsit-dms}^{f}$ satisfies the secrecy constraint in (\ref{wawa2}).
In the preceding subsection, we introduce an extended feedback scheme for the GMAC-NCSIT-DMS with feedback,
and show that this scheme achieves $\mathcal{C}_{gmac-ncsit-dms}^{f}$.
In this new scheme, the transmitted codewords $X_{1,i}$, $U_{i}$ and $V_{i}$ at time $i$ ($1\leq i\leq N$) can be expressed
almost in the same fashion as those in (\ref{gxt1}), except that
\begin{eqnarray}\label{xxs.gxt1}
&&V_{1}=\sqrt{12(1-\rho^{2})P_{2}}(\theta_{2}-\frac{S_{1}}{\sqrt{12(1-\rho^{2})P_{2}}}+A_{2}),\nonumber\\
&&X_{1,2}=\frac{\sqrt{12P^{*}}(\theta_{1}-\frac{S_{2}}{\sqrt{12P^{*}}}+A_{1})}{\rho\sqrt{\frac{P_{2}}{P_{1}}}+1},\,\,U_{2}=\rho\sqrt{\frac{P_{2}}{P_{1}}}X_{1,2}.
\end{eqnarray}
From (\ref{gxt1}) and (\ref{xxs.gxt1}), we can conclude that for $3\leq i\leq N$, $\theta_{1}$ and $\theta_{2}$ are not contained in
the transmitted $X_{1,i}$, $U_{i}$ and $V_{i}$. Hence following the steps in (\ref{quanyou2}) and choosing sufficiently large $N$, we can prove that
$\frac{1}{N}H(W_{1},W_{2}|Z^{N})\geq R_{1}+R_{2}-\epsilon$, which completes the proof.
\end{IEEEproof}

For comparison, the following Corollary \ref{T9} establishes
an outer bound on the secrecy capacity region $\mathcal{C}_{s,gmac-ncsit-dms}$ of GMAC-WT-NCSIT-DMS.

\begin{corollary}\label{T9}
$\mathcal{C}_{s,gmac-ncsit-dms}\subseteq\mathcal{C}^{out}_{s,gmac-ncsit-dms}$, where $\mathcal{C}^{out}_{s,gmac-ncsit-dms}$ is given by
\begin{eqnarray}\label{gg.xxs.liqin1}
&&\mathcal{C}^{out}_{s,gmac-ncsit-dms}=\bigcup_{-1\leq\rho_{12},\rho_{1s},\rho_{2s}\leq 1}\left\{(R_{1}\geq 0,R_{2}\geq 0):\right.\nonumber\\
&&R_{2}\leq\frac{1}{2}\log\left(1+\frac{P_{2}+\sigma_{1}^{2}+a^{2}P_{1}+b^{2}Q-2a\rho_{12}\sqrt{P_{1}P_{2}}-2b\rho_{2s}\sqrt{P_{2}Q}
+2ab\rho_{1s}\sqrt{P_{1}Q}}{\sigma_{1}^{2}}\right),\nonumber\\
&&\left.R_{1}+R_{2}\leq \frac{1}{2}\log\left(1+\frac{P_{1}+P_{2}+Q+2\sqrt{P_{1}P_{2}}\rho_{12}+2\rho_{1s}\sqrt{P_{1}Q}+2\rho_{2s}\sqrt{P_{2}Q}}{\sigma_{1}^{2}}\right)\right.\nonumber\\
&&\left.-\frac{1}{2}\log\left(1+\frac{P_{1}+P_{2}+Q+2\sqrt{P_{1}P_{2}}\rho_{12}+2\rho_{1s}\sqrt{P_{1}Q}+2\rho_{2s}\sqrt{P_{2}Q}}{\sigma_{1}^{2}+\sigma_{2}^{2}}\right)\right\},\nonumber\\
\end{eqnarray}
where
\begin{eqnarray}\label{gg.xxs.liqin1.hh}
&&a=\sqrt{\frac{P_{2}}{P_{1}}}\frac{\rho_{12}-\rho_{1s}\rho_{2s}}{1-\rho^{2}_{1s}},\,\,\,b=\sqrt{\frac{P_{2}}{Q}}\frac{\rho_{2s}-\rho_{12}\rho_{1s}}{1-\rho^{2}_{1s}}.
\end{eqnarray}
\end{corollary}
\begin{IEEEproof}
See Appendix \ref{appen2}.
\end{IEEEproof}

The following Figure \ref{f1x} shows the rate gains by using channel feedback for
$P_{1}=10$, $P_{2}=3$, $Q=5$, $\sigma^{2}_{1}=10$ and $\sigma_{2}^{2}=20$. 

\begin{figure}[htb]
\centering
\includegraphics[scale=0.5]{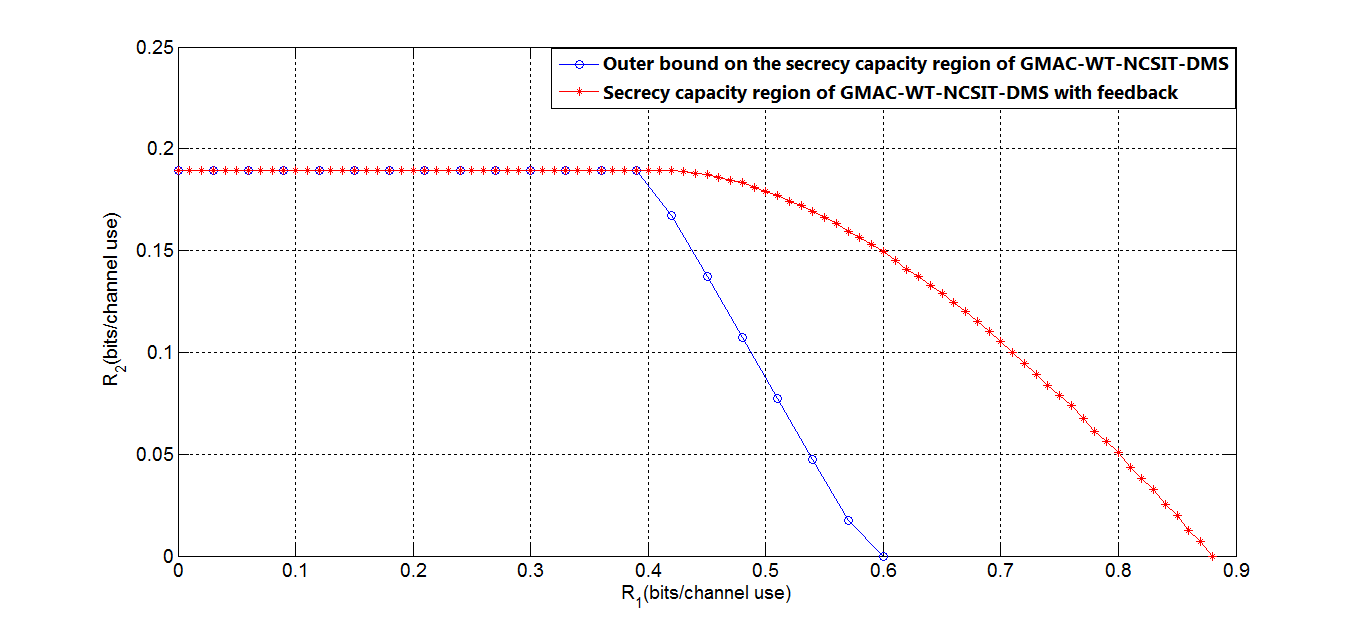}
\caption{Capacity results on GMAC-WT-NCSIT-DMS with or without feedback}
\label{f1x}
\end{figure}

\section{Conclusion\label{secVI}}
\setcounter{equation}{0}

In this paper, we determine the secrecy capacity regions of the GMAC-WT-DMS with feedback and the GMAC-WT-NCSIT-DMS with feedback by
proposing SSCA feedback schemes for these models.
Possible future work includes:
\begin{itemize}

\item The rate-splitting feature used in \cite{tangxiaojun} might be a good element to identify future
strategies that potentially be useful for the GMACs with general (not necessarily degraded) message
set, and maybe via that one can reach similar conclusions given above when the rate regions are not
degraded by introducing secrecy constraint.

\item To explore
whether one can identify dualities of some kind between the GMAC and the Gaussian broadcast models
when feedback and secrecy constraint are considered.

\item The finite blocklength regime also deserves attention
even in the single user wiretap case where a modified SK
scheme motivated by \cite{c3} might be useful.

\end{itemize}

\section*{Acknowledgment}
The work of B. Dai was supported by
the National Natural Science Foundation of China under Grant 62071392,
and the 111 Project No.111-2-14. The work of Y. Liang was supported by U.S. NSF CCF-1801846.
The work of S. Shamai was supported by
the European Union's Horizon 2020 Research And Innovation Programme under Grant 694630.

\renewcommand{\theequation}{A\arabic{equation}}
\setcounter{equation}{0}
\appendices\section{Proof of Lemma \ref{L1}\label{appenx}}

For $2\leq k\leq N$, define
\begin{eqnarray}\label{qqe1.xxs11}
&&\eta^{'}_{1,k}=\eta_{1,k}+V_{k}.
\end{eqnarray}
Note that
\begin{eqnarray}\label{qqe1.xxs12}
E[(\eta^{'}_{1,k})^{2}]&=&E[(\eta_{1,k}+V_{k})^{2}]\nonumber\\
&\stackrel{(a)}=&E[(\eta_{1,k})^{2}]+E[(V_{k})^{2}]\stackrel{(b)}=\sigma_{1}^{2}+(1-\rho^{2})P_{2},
\end{eqnarray}
where (a) follows from the fact that $V_{k}$ is independent of $\eta_{1,k}$ since $V_{1}$ is a function of $\theta_{1}$ and
$V_{k}$ ($2\leq k\leq N$) is a function of $\eta_{1,1}$,...,$\eta_{1,k-1}$, and (b) follows from
(\ref{qqe1.xxs4}). Furthermore, from (\ref{qqe1.xxs3}) and (\ref{qqe1.xxs4}), $V_{k}$ can be re-written as
\begin{eqnarray}\label{qqe1.xxs13}
V_{k}&=&\sqrt{\frac{(1-\rho^{2})P_{2}}{\alpha_{k-1}}}\epsilon_{k-1}\nonumber\\
&=&\sqrt{\frac{(1-\rho^{2})P_{2}}{\alpha_{k-1}}}\left(\epsilon_{k-2}-\frac{E[(V_{k-1}+\eta_{1,k-1})\epsilon_{k-2}]}{E[(V_{k-1}+\eta_{1,k-1})^{2}]}(V_{k-1}+\eta_{1,k-1})\right)\nonumber\\
&\stackrel{(c)}=&\sqrt{\frac{(1-\rho^{2})P_{2}}{\alpha_{k-1}}}\left(\epsilon_{k-2}-\frac{\sqrt{(1-\rho^{2})P_{2}\alpha_{k-2}}}{(1-\rho^{2})P_{2}+\sigma_{1}^{2}}(V_{k-1}+\eta_{1,k-1})\right)\nonumber\\
&=&\sqrt{\frac{(1-\rho^{2})P_{2}}{\alpha_{k-2}}}\sqrt{\frac{\alpha_{k-2}}{\alpha_{k-1}}}\left(\epsilon_{k-2}-\frac{\sqrt{(1-\rho^{2})P_{2}\alpha_{k-2}}}{(1-\rho^{2})P_{2}+\sigma_{1}^{2}}(V_{k-1}+\eta_{1,k-1})\right)\nonumber\\
&\stackrel{(d)}=&\sqrt{\frac{\alpha_{k-2}}{\alpha_{k-1}}}V_{k-1}-\sqrt{\frac{(1-\rho^{2})P_{2}}{\alpha_{k-1}}}\frac{\sqrt{(1-\rho^{2})P_{2}\alpha_{k-2}}}{(1-\rho^{2})P_{2}+\sigma_{1}^{2}}(V_{k-1}+\eta_{1,k-1})\nonumber\\
&=&\sqrt{\frac{\alpha_{k-2}}{\alpha_{k-1}}}\frac{\sigma_{1}^{2}}{(1-\rho^{2})P_{2}+\sigma_{1}^{2}}V_{k-1}-\sqrt{\frac{\alpha_{k-2}}{\alpha_{k-1}}}\frac{(1-\rho^{2})P_{2}}{(1-\rho^{2})P_{2}+\sigma_{1}^{2}}\eta_{1,k-1},
\end{eqnarray}
where (c) follows from $\epsilon_{k-2}$ is independent of $\eta_{1,k-1}$, $V_{k-1}=\sqrt{\frac{(1-\rho^{2})P_{2}}{\alpha_{k-2}}}\epsilon_{k-2}$ and
$\alpha_{k-2}\triangleq Var(\epsilon_{k-2})$, and (d) follows from $V_{k-1}=\sqrt{\frac{(1-\rho^{2})P_{2}}{\alpha_{k-2}}}\epsilon_{k-2}$.
Substituting (\ref{qqe1.xxs13}) into (\ref{qqe1.xxs11}), we have
\begin{eqnarray}\label{qqe1.xxs14}
\eta^{'}_{1,k}&=&\eta_{1,k}+V_{k}=\eta_{1,k}+\sqrt{\frac{\alpha_{k-2}}{\alpha_{k-1}}}\frac{\sigma_{1}^{2}}{(1-\rho^{2})P_{2}+\sigma_{1}^{2}}V_{k-1}
-\sqrt{\frac{\alpha_{k-2}}{\alpha_{k-1}}}\frac{(1-\rho^{2})P_{2}}{(1-\rho^{2})P_{2}+\sigma_{1}^{2}}\eta_{1,k-1}\nonumber\\
&=&\eta_{1,k}+\sqrt{\frac{\alpha_{k-2}}{\alpha_{k-1}}}\frac{\sigma_{1}^{2}}{(1-\rho^{2})P_{2}+\sigma_{1}^{2}}V_{k-1}
-\sqrt{\frac{\alpha_{k-2}}{\alpha_{k-1}}}\frac{(1-\rho^{2})P_{2}}{(1-\rho^{2})P_{2}+\sigma_{1}^{2}}\eta_{1,k-1}\nonumber\\
&&+\sqrt{\frac{\alpha_{k-2}}{\alpha_{k-1}}}\frac{\sigma_{1}^{2}}{(1-\rho^{2})P_{2}+\sigma_{1}^{2}}\eta_{1,k-1}
-\sqrt{\frac{\alpha_{k-2}}{\alpha_{k-1}}}\frac{\sigma_{1}^{2}}{(1-\rho^{2})P_{2}+\sigma_{1}^{2}}\eta_{1,k-1}\nonumber\\
&=&\eta_{1,k}+\sqrt{\frac{\alpha_{k-2}}{\alpha_{k-1}}}\frac{\sigma_{1}^{2}}{(1-\rho^{2})P_{2}+\sigma_{1}^{2}}(V_{k-1}+\eta_{1,k-1})
-\sqrt{\frac{\alpha_{k-2}}{\alpha_{k-1}}}\eta_{1,k-1}\nonumber\\
&=&\eta_{1,k}+\sqrt{\frac{\alpha_{k-2}}{\alpha_{k-1}}}\frac{\sigma_{1}^{2}}{(1-\rho^{2})P_{2}+\sigma_{1}^{2}}\eta^{'}_{1,k-1}
-\sqrt{\frac{\alpha_{k-2}}{\alpha_{k-1}}}\eta_{1,k-1}.
\end{eqnarray}
From classical SK scheme \cite{sk}, we know that
\begin{eqnarray}\label{qqe1.xxs15}
&&\frac{\alpha_{k}}{\alpha_{k-1}}=\frac{\sigma_{1}^{2}}{(1-\rho^{2})P_{2}+\sigma_{1}^{2}}
\end{eqnarray}
for all $2\leq k\leq N$.
Substituting (\ref{qqe1.xxs15}) into (\ref{qqe1.xxs14}), we obtain
\begin{eqnarray}\label{qqe1.xxs16}
&&\eta^{'}_{1,k}=\frac{\sigma_{1}}{\sqrt{(1-\rho^{2})P_{2}+\sigma_{1}^{2}}}\eta^{'}_{1,k-1}+\eta_{1,k}
-\sqrt{\frac{(1-\rho^{2})P_{2}+\sigma_{1}^{2}}{\sigma_{1}^{2}}}\eta_{1,k-1}.
\end{eqnarray}
Here note that (\ref{qqe1.xxs16}) holds for $3\leq k\leq N$, and
\begin{eqnarray}\label{sheva1.1}
&&\eta^{'}_{1,2}=\eta_{1,2}+V_{2}=\eta_{1,2}+\sqrt{\frac{(1-\rho^{2})P_{2}}{\alpha_{1}}}\epsilon_{1}
=\eta_{1,2}+\frac{\eta_{1,1}\sqrt{(1-\rho^{2})P_{2}}}{\sigma_{1}}.
\end{eqnarray}
On the other hand, from (\ref{qqe1.xxs6}), we have
\begin{eqnarray}\label{qqe1.xxs17}
&&E[Y_{k-1}\epsilon^{'}_{k-2}]=E[(X^{*}_{k-1}+\eta^{'}_{1,k-1})\epsilon^{'}_{k-2}]\nonumber\\
&&\stackrel{(e)}=E\left[\left(\sqrt{\frac{P^{*}}{\alpha^{'}_{k-2}}}\epsilon^{'}_{k-2}+\eta^{'}_{1,k-1}\right)\epsilon^{'}_{k-2}\right]
=\sqrt{P^{*}\alpha^{'}_{k-2}}+E[\eta^{'}_{1,k-1}\epsilon^{'}_{k-2}],
\end{eqnarray}
and
\begin{eqnarray}\label{qqe1.xxs18}
&&E[Y^{2}_{k-1}]=E[(X^{*}_{k-1}+\eta^{'}_{1,k-1})^{2}]=E\left[\left(\sqrt{\frac{P^{*}}{\alpha^{'}_{k-2}}}\epsilon^{'}_{k-2}+\eta^{'}_{1,k-1}\right)^{2}\right]\nonumber\\
&&\stackrel{(f)}=P^{*}+2\sqrt{\frac{P^{*}}{\alpha^{'}_{k-2}}}E[\epsilon^{'}_{k-2}\eta^{'}_{1,k-1}]+(1-\rho^{2})P_{2}+\sigma_{1}^{2},
\end{eqnarray}
where (e) follows from (\ref{qqe1.xxs5}), and (f) follows from (\ref{qqe1.xxs12}). Substituting (\ref{qqe1.xxs17}) and (\ref{qqe1.xxs18})
into (\ref{qqe1.xxs6}), $\epsilon^{'}_{k-1}$ can be re-written as
\begin{eqnarray}\label{qqe1.xxs19}
&&\epsilon^{'}_{k-1}=\epsilon^{'}_{k-2}-\frac{E[Y_{k-1}\epsilon^{'}_{k-2}]}{E[Y^{2}_{k-1}]}Y_{k-1}\nonumber\\
&&=\epsilon^{'}_{k-2}-\frac{\sqrt{P^{*}\alpha^{'}_{k-2}}+E[\epsilon^{'}_{k-2}\eta^{'}_{1,k-1}]}{P^{*}+2\sqrt{\frac{P^{*}}{\alpha^{'}_{k-2}}}E[\epsilon^{'}_{k-2}\eta^{'}_{1,k-1}]+(1-\rho^{2})P_{2}+\sigma_{1}^{2}}
\left(\sqrt{\frac{P^{*}}{\alpha^{'}_{k-2}}}\epsilon^{'}_{k-2}+\eta^{'}_{1,k-1}\right)\nonumber\\
&&=\epsilon^{'}_{k-2}-\frac{\epsilon^{'}_{k-2}(P^{*}+E[\epsilon^{'}_{k-2}\eta^{'}_{1,k-1}]\sqrt{\frac{P^{*}}{\alpha^{'}_{k-2}}})
+\eta^{'}_{1,k-1}(\sqrt{P^{*}\cdot\alpha^{'}_{k-2}}+E[\epsilon^{'}_{k-2}\eta^{'}_{1,k-1}])}{P^{*}+2\sqrt{\frac{P^{*}}{\alpha^{'}_{k-2}}}E[\epsilon^{'}_{k-2}\eta^{'}_{1,k-1}]+(1-\rho^{2})P_{2}+\sigma_{1}^{2}}\nonumber\\
&&=\epsilon^{'}_{k-2}\frac{\sqrt{\frac{P^{*}}{\alpha^{'}_{k-2}}}E[\epsilon^{'}_{k-2}\eta^{'}_{1,k-1}]+(1-\rho^{2})P_{2}+\sigma_{1}^{2}}{P^{*}
+2\sqrt{\frac{P^{*}}{\alpha^{'}_{k-2}}}E[\epsilon^{'}_{k-2}\eta^{'}_{1,k-1}]+(1-\rho^{2})P_{2}+\sigma_{1}^{2}}\nonumber\\
&&-\eta^{'}_{1,k-1}\frac{\sqrt{P^{*}\cdot\alpha^{'}_{k-2}}+E[\epsilon^{'}_{k-2}\eta^{'}_{1,k-1}]}{P^{*}+2\sqrt{\frac{P^{*}}{\alpha^{'}_{k-2}}}
E[\epsilon^{'}_{k-2}\eta^{'}_{1,k-1}]+(1-\rho^{2})P_{2}+\sigma_{1}^{2}}.
\end{eqnarray}
From (\ref{qqe1.xxs19}), we observe that $\epsilon^{'}_{k-1}$ depends on $E[\epsilon^{'}_{k-2}\eta^{'}_{1,k-1}]$.
Combining (\ref{qqe1.xxs16}) with (\ref{qqe1.xxs19}), we can conclude that
\begin{eqnarray}\label{qqe1.xxs20}
&&E[\epsilon^{'}_{k-1}\eta^{'}_{1,k}]=E[(\frac{\sigma_{1}}{\sqrt{(1-\rho^{2})P_{2}+\sigma_{1}^{2}}}\eta^{'}_{1,k-1}+\eta_{1,k}
-\sqrt{\frac{(1-\rho^{2})P_{2}+\sigma_{1}^{2}}{\sigma_{1}^{2}}}\eta_{1,k-1})\nonumber\\
&&\cdot(\epsilon^{'}_{k-2}\frac{\sqrt{\frac{P^{*}}{\alpha^{'}_{k-2}}}
E[\epsilon^{'}_{k-2}\eta^{'}_{1,k-1}]+(1-\rho^{2})P_{2}+\sigma_{1}^{2}}{P^{*}+2\sqrt{\frac{P^{*}}{\alpha^{'}_{k-2}}}
E[\epsilon^{'}_{k-2}\eta^{'}_{1,k-1}]+(1-\rho^{2})P_{2}+\sigma_{1}^{2}}\nonumber\\
&&-\eta^{'}_{1,k-1}\frac{\sqrt{P^{*}\cdot\alpha^{'}_{k-2}}+E[\epsilon^{'}_{k-2}\eta^{'}_{1,k-1}]}
{P^{*}+2\sqrt{\frac{P^{*}}{\alpha^{'}_{k-2}}}E[\epsilon^{'}_{k-2}\eta^{'}_{1,k-1}]+(1-\rho^{2})P_{2}+\sigma_{1}^{2}})]\nonumber\\
&&\stackrel{(g)}=\frac{\sqrt{\frac{P^{*}}{\alpha^{'}_{k-2}}}
E[\epsilon^{'}_{k-2}\eta^{'}_{1,k-1}]+(1-\rho^{2})P_{2}+\sigma_{1}^{2}}{P^{*}+2\sqrt{\frac{P^{*}}{\alpha^{'}_{k-2}}}
E[\epsilon^{'}_{k-2}\eta^{'}_{1,k-1}]+(1-\rho^{2})P_{2}+\sigma_{1}^{2}}\cdot\frac{\sigma_{1}}
{\sqrt{(1-\rho^{2})P_{2}+\sigma_{1}^{2}}}E[\epsilon^{'}_{k-2}\eta^{'}_{1,k-1}]\nonumber\\
&&-\frac{\sqrt{P^{*}\cdot\alpha^{'}_{k-2}}+E[\epsilon^{'}_{k-2}\eta^{'}_{1,k-1}]}
{P^{*}+2\sqrt{\frac{P^{*}}{\alpha^{'}_{k-2}}}E[\epsilon^{'}_{k-2}\eta^{'}_{1,k-1}]+(1-\rho^{2})P_{2}+\sigma_{1}^{2}}\cdot\frac{\sigma_{1}}{\sqrt{(1-\rho^{2})P_{2}+\sigma_{1}^{2}}}
E[(\eta^{'}_{1,k-1})^{2}]\nonumber\\
&&+\frac{\sqrt{P^{*}\cdot\alpha^{'}_{k-2}}+E[\epsilon^{'}_{k-2}\eta^{'}_{1,k-1}]}
{P^{*}+2\sqrt{\frac{P^{*}}{\alpha^{'}_{k-2}}}E[\epsilon^{'}_{k-2}\eta^{'}_{1,k-1}]+(1-\rho^{2})P_{2}+\sigma_{1}^{2}}\cdot
\sqrt{\frac{(1-\rho^{2})P_{2}+\sigma_{1}^{2}}{\sigma_{1}^{2}}}E[\eta_{1,k-1}\eta^{'}_{1,k-1}]\nonumber\\
&&\stackrel{(h)}=\frac{\sqrt{\frac{P^{*}}{\alpha^{'}_{k-2}}}
E[\epsilon^{'}_{k-2}\eta^{'}_{1,k-1}]+(1-\rho^{2})P_{2}+\sigma_{1}^{2}}{P^{*}+2\sqrt{\frac{P^{*}}{\alpha^{'}_{k-2}}}
E[\epsilon^{'}_{k-2}\eta^{'}_{1,k-1}]+(1-\rho^{2})P_{2}+\sigma_{1}^{2}}\cdot\frac{\sigma_{1}}
{\sqrt{(1-\rho^{2})P_{2}+\sigma_{1}^{2}}}E[\epsilon^{'}_{k-2}\eta^{'}_{1,k-1}]\nonumber\\
&&-\frac{\sqrt{P^{*}\cdot\alpha^{'}_{k-2}}+E[\epsilon^{'}_{k-2}\eta^{'}_{1,k-1}]}
{P^{*}+2\sqrt{\frac{P^{*}}{\alpha^{'}_{k-2}}}E[\epsilon^{'}_{k-2}\eta^{'}_{1,k-1}]+(1-\rho^{2})P_{2}+\sigma_{1}^{2}}\cdot
\sigma_{1}\sqrt{(1-\rho^{2})P_{2}+\sigma_{1}^{2}}\nonumber\\
&&+\frac{\sqrt{P^{*}\cdot\alpha^{'}_{k-2}}+E[\epsilon^{'}_{k-2}\eta^{'}_{1,k-1}]}
{P^{*}+2\sqrt{\frac{P^{*}}{\alpha^{'}_{k-2}}}E[\epsilon^{'}_{k-2}\eta^{'}_{1,k-1}]+(1-\rho^{2})P_{2}+\sigma_{1}^{2}}\cdot
\sigma_{1}\sqrt{(1-\rho^{2})P_{2}+\sigma_{1}^{2}}\nonumber\\
&&=\frac{\sqrt{\frac{P^{*}}{\alpha^{'}_{k-2}}}
E[\epsilon^{'}_{k-2}\eta^{'}_{1,k-1}]+(1-\rho^{2})P_{2}+\sigma_{1}^{2}}{P^{*}+2\sqrt{\frac{P^{*}}{\alpha^{'}_{k-2}}}
E[\epsilon^{'}_{k-2}\eta^{'}_{1,k-1}]+(1-\rho^{2})P_{2}+\sigma_{1}^{2}}\cdot\frac{\sigma_{1}}
{\sqrt{(1-\rho^{2})P_{2}+\sigma_{1}^{2}}}E[\epsilon^{'}_{k-2}\eta^{'}_{1,k-1}],\nonumber\\
\end{eqnarray}
where (g) follows from $E[\epsilon^{'}_{k-2}\eta_{1,k}]=E[\epsilon^{'}_{k-2}\eta_{1,k-1}]=E[\eta^{'}_{1,k-1}\eta_{1,k}]=0$,
and (h) follows from (\ref{qqe1.xxs16}), which indicates that
\begin{eqnarray}\label{qqe1.xxs21}
E[\eta^{'}_{1,k-1}\eta_{1,k-1}]&=&E\left[\left(\frac{\sigma_{1}}{\sqrt{(1-\rho^{2})P_{2}+\sigma_{1}^{2}}}\eta^{'}_{1,k-2}+\eta_{1,k-1}
-\sqrt{\frac{(1-\rho^{2})P_{2}+\sigma_{1}^{2}}{\sigma_{1}^{2}}}\eta_{1,k-2}\right)\eta_{1,k-1}\right]\nonumber\\
&\stackrel{(i)}=&E[(\eta_{1,k-1})^{2}]=\sigma_{1}^{2},
\end{eqnarray}
where (i) follows from $E[\eta^{'}_{1,k-2}\eta_{1,k-1}]=E[\eta_{1,k-2}\eta_{1,k-1}]=0$.

Observing that the first item of $E[\epsilon^{'}_{k-1}\eta^{'}_{1,k}]$ is $E[\epsilon^{'}_{2}\eta^{'}_{1,3}]$, and it is given by
\begin{eqnarray}\label{qqe1.xxs22}
&&E[\epsilon^{'}_{2}\eta^{'}_{1,3}]=E[\epsilon^{'}_{2}(V_{3}+\eta_{1,3})]=E\left[\epsilon^{'}_{2}(\eta_{1,3}+\sqrt{\frac{(1-\rho^{2})P_{2}}{\alpha_{2}}}\epsilon_{2})\right]\nonumber\\
&&\stackrel{(j)}=E\left[\frac{\sqrt{\frac{(1-\rho^{2})P_{2}}{\sigma_{1}^{2}}}\eta_{1,1}+\eta_{1,2}}{\sqrt{12P^{*}}}
(\eta_{1,3}+\frac{\sqrt{(1-\rho^{2})P_{2}}}{r}\eta_{1,1}-\frac{(1-\rho^{2})P_{2}}{r\sigma_{1}}\eta_{1,2})\right]\nonumber\\
&&\stackrel{(k)}=\frac{(1-\rho^{2})P_{2}}{r\sigma_{1}}\frac{\sigma_{1}^{2}}{\sqrt{12P^{*}}}-\frac{(1-\rho^{2})P_{2}}{r\sigma_{1}}\frac{\sigma_{1}^{2}}{\sqrt{12P^{*}}}=0,
\end{eqnarray}
where (j) follows from
\begin{eqnarray}\label{sheva.x1}
&&\epsilon^{'}_{2}=\frac{V_{2}+\eta_{1,2}}{\sqrt{12P^{*}}}=\frac{\sqrt{\frac{(1-\rho^{2})P_{2}}{\sigma_{1}^{2}}}\eta_{1,1}+\eta_{1,2}}{\sqrt{12P^{*}}},
\end{eqnarray}
\begin{eqnarray}\label{sheva.x2}
&&\epsilon_{2}=\epsilon_{1}-\frac{E[Y_{2}^{'}\epsilon_{1}]}{E[Y^{'2}_{2}]}Y^{'}_{2}\nonumber\\
&&=\frac{\eta_{1,1}}{\sqrt{12(1-\rho^{2})P_{2}}}-\frac{E[(\sqrt{\frac{(1-\rho^{2})P_{2}}{\sigma_{1}^{2}}}\eta_{1,1}+\eta_{1,2})\frac{\eta_{1,1}}{\sqrt{12(1-\rho^{2})P_{2}}}]}
{E[(\sqrt{\frac{(1-\rho^{2})P_{2}}{\sigma_{1}^{2}}}\eta_{1,1}+\eta_{1,2})^{2}]}(\sqrt{\frac{(1-\rho^{2})P_{2}}{\sigma_{1}^{2}}}\eta_{1,1}+\eta_{1,2})\nonumber\\
&&=\frac{\sigma_{1}^{2}}{\sqrt{12(1-\rho^{2})P_{2}}r^{2}}\eta_{1,1}-\frac{\sigma_{1}}{\sqrt{12}r^{2}}\eta_{1,2},
\end{eqnarray}
\begin{eqnarray}\label{sheva.x3}
&&\alpha_{2}=\frac{\sigma_{1}^{4}}{12(1-\rho^{2})P_{2}r^{2}},
\end{eqnarray}
\begin{eqnarray}\label{qqe1.xxs24}
&&r=\sqrt{(1-\rho^{2})P_{2}+\sigma_{1}^{2}},
\end{eqnarray}
and (k) follows from $E[\eta_{1,3}\eta_{1,1}]=E[\eta_{1,3}\eta_{1,2}]=E[\eta_{1,1}\eta_{1,2}]=0$.
Now substituting (\ref{qqe1.xxs22}) into (\ref{qqe1.xxs20}), we can conclude that
\begin{eqnarray}\label{dlrb1}
&&E[\epsilon^{'}_{k-1}\eta^{'}_{1,k}]=0
\end{eqnarray}
for all $3\leq k\leq N$, which completes the proof.

\section{Proof of Corollary \ref{T6}\label{appen1}}

We begin with the sum rate bound $R_{1}+R_{2}-\frac{1}{N}H(W_{1},W_{2}|Z^{N})\leq \epsilon$, which is bounded by 
\begin{eqnarray}\label{xiaobao1}
&&R_{1}+R_{2}-\epsilon\leq \frac{1}{N}H(W_{1},W_{2}|Z^{N})\nonumber\\
&&=\frac{1}{N}(H(W_{1},W_{2}|Z^{N})
-H(W_{1},W_{2}|Z^{N},Y^{N})+H(W_{1},W_{2}|Z^{N},Y^{N}))\nonumber\\
&&\stackrel{(b)}\leq \frac{1}{N}(I(W_{1},W_{2};Y^{N}|Z^{N})+\delta(\epsilon))\nonumber\\
&&\stackrel{(c)}\leq \frac{1}{N}(I(X_{1}^{N},X_{2}^{N};Y^{N}|Z^{N})+\delta(\epsilon))\nonumber\\
&&\stackrel{(d)}=\frac{1}{N}(I(X_{1}^{N},X_{2}^{N};Y^{N})-I(X_{1}^{N},X_{2}^{N};Z^{N})+\delta(\epsilon))\nonumber\\
&&=\frac{1}{N}\sum_{i=1}^{N}(H(Y_{i}|Y^{i-1})-H(Y_{i}|X_{1,i},X_{2,i})-H(Z_{i}|Z^{i-1})+H(Z_{i}|X_{1,i},X_{2,i}))+\frac{\delta(\epsilon)}{N}\nonumber\\
&&\stackrel{(e)}=\frac{1}{N}\sum_{i=1}^{N}(H(Y_{i}|Y^{i-1},Z^{i-1})-H(Y_{i}|X_{1,i},X_{2,i})-H(Z_{i}|Z^{i-1})+H(Z_{i}|X_{1,i},X_{2,i}))+\frac{\delta(\epsilon)}{N}\nonumber\\
&&\leq \frac{1}{N}\sum_{i=1}^{N}(H(Y_{i}|Z^{i-1})-H(Y_{i}|X_{1,i},X_{2,i})-H(Z_{i}|Z^{i-1})+H(Z_{i}|X_{1,i},X_{2,i}))+\frac{\delta(\epsilon)}{N}\nonumber\\
&&\stackrel{(f)}\leq \frac{1}{N}\sum_{i=1}^{N}(H(Y_{i})-H(Y_{i}|X_{1,i},X_{2,i})-H(Z_{i})+H(Z_{i}|X_{1,i},X_{2,i}))+\frac{\delta(\epsilon)}{N}\nonumber\\
&&=\frac{1}{N}\sum_{i=1}^{N}(I(X_{1,i},X_{2,i};Y_{i})-I(X_{1,i},X_{2,i};Z_{i}))+\frac{\delta(\epsilon)}{N},
\end{eqnarray}
where (b) follows from Fano's inequality and $P_{e}\leq \epsilon$, (c) follows from $H(W_{1},W_{2}|X_{1}^{N},X_{2}^{N})=0$,
(d) follows from
$(X_{1}^{N},X_{2}^{N})\rightarrow Y^{N}\rightarrow Z^{N}$, (e) follows from $Y_{i}\rightarrow Y^{i-1}\rightarrow Z^{i-1}$, and
(f) follows from $Z^{i-1}\rightarrow Y_{i}\rightarrow Z_{i}$, which indicates that $I(Z_{i};Z^{i-1})\leq I(Y_{i};Z^{i-1})$, i.e.,
$H(Y_{i}|Z^{i-1})-H(Z_{i}|Z^{i-1})\leq H(Y_{i})-H(Z_{i})$.

Then substituting $Y_{i}=X_{1,i}+X_{2,i}+\eta_{1,i}$ and $Z_{i}=Y_{i}+\eta_{2,i}$ into (\ref{xiaobao1}),
we have
\begin{eqnarray}\label{xiaobao3}
&&R_{1}+R_{2}-\epsilon\leq\frac{1}{N}\sum_{i=1}^{N}h(Y_{i})-\frac{1}{N}\sum_{i=1}^{N}h(Y_{i}|X_{1,i},X_{2,i})
-\frac{1}{N}\sum_{i=1}^{N}h(Z_{i})+\frac{1}{N}\sum_{i=1}^{N}h(Z_{i}|X_{1,i},X_{2,i})+\frac{\delta(\epsilon)}{N}\nonumber\\
&&\stackrel{(a)}=\frac{1}{N}\sum_{i=1}^{N}h(Y_{i})-\frac{1}{N}\sum_{i=1}^{N}h(\eta_{1,i})
-\frac{1}{N}\sum_{i=1}^{N}h(Z_{i})+\frac{1}{N}\sum_{i=1}^{N}h(\eta_{1,i}+\eta_{2,i})+\frac{\delta(\epsilon)}{N}\nonumber\\
&&\stackrel{(b)}\leq\frac{1}{N}\sum_{i=1}^{N}h(Y_{i})-\frac{1}{2}\log2\pi e\sigma_{1}^{2}
-\frac{1}{N}\sum_{i=1}^{N}\frac{1}{2}\log(2^{2h(Y_{i})}+2^{2h(\eta_{2,i})})+\frac{1}{2}\log2\pi e(\sigma_{1}^{2}+\sigma_{2}^{2})+\frac{\delta(\epsilon)}{N}\nonumber\\
&&\stackrel{(c)}\leq\frac{1}{N}\sum_{i=1}^{N}h(Y_{i})-\frac{1}{2}\log2\pi e\sigma_{1}^{2}
-\frac{1}{2}\log(2^{2\frac{1}{N}\sum_{i=1}^{N}h(Y_{i})}+2\pi e\sigma_{2}^{2})+\frac{1}{2}\log2\pi e(\sigma_{1}^{2}+\sigma_{2}^{2})+\frac{\delta(\epsilon)}{N}\nonumber\\
&&\stackrel{(d)}\leq\frac{1}{2}\log2\pi e(P_{1}+P_{2}+2\sqrt{P_{1}P_{2}}\rho+\sigma_{1}^{2})-\frac{1}{2}\log2\pi e\sigma_{1}^{2}
-\frac{1}{2}\log(2\pi e(P_{1}+P_{2}+2\sqrt{P_{1}P_{2}}\rho+\sigma_{1}^{2}+\sigma_{2}^{2}))\nonumber\\
&&+\frac{1}{2}\log2\pi e(\sigma_{1}^{2}+\sigma_{2}^{2})+\frac{\delta(\epsilon)}{N}\nonumber\\
&&=\frac{1}{2}\log(1+\frac{P_{1}+P_{2}+2\sqrt{P_{1}P_{2}}\rho}{\sigma_{1}^{2}})
-\frac{1}{2}\log(1+\frac{P_{1}+P_{2}+2\sqrt{P_{1}P_{2}}\rho}{\sigma_{1}^{2}+\sigma_{2}^{2}})+\frac{\delta(\epsilon)}{N},
\end{eqnarray}
where (a) follows from $\eta_{1,i}$ and $\eta_{2,i}$ are independent of $X_{1,i}$ and $X_{2,i}$, (b) follows from
the entropy power inequality, (c) follows from the fact that $\log(2^{x}+c)$ is a convex function and Jensen's inequality,
and (d) follows from $\frac{1}{N}\sum_{i=1}^{N}h(Y_{i})-\frac{1}{2}\log(2^{2\frac{1}{N}\sum_{i=1}^{N}h(Y_{i})}+2\pi e\sigma_{2}^{2})$
is increasing while $\frac{1}{N}\sum_{i=1}^{N}h(Y_{i})$ is increasing and
\begin{eqnarray}\label{xiaobao4}
&&\frac{1}{N}\sum_{i=1}^{N}h(Y_{i})\nonumber\\
&&\leq \frac{1}{N}\sum_{i=1}^{N}\frac{1}{2}\log2\pi e(P_{1,i}+P_{2,i}+2E[X_{1,i}X_{2,i}]+\sigma_{1}^{2})\nonumber\\
&&\leq\frac{1}{2}\log2\pi e(\frac{1}{N}\sum_{i=1}^{N}(P_{1,i}+P_{2,i}+2E[X_{1,i}X_{2,i}])+\sigma_{1}^{2})
\stackrel{(e)}=\frac{1}{2}\log2\pi e(P_{1}+P_{2}+2\sqrt{P_{1}P_{2}}\rho+\sigma_{1}^{2}),\nonumber\\
\end{eqnarray}
where (e) follows from the definitions
\begin{eqnarray}\label{xiaobao2}
&&E[X^{2}_{1,i}]=P_{1,i},\,\,E[X^{2}_{2,i}]=P_{2,i},\,\,P_{1}=\frac{1}{N}\sum_{i=1}^{N}P_{1,i},\,\,P_{2}=\frac{1}{N}\sum_{i=1}^{N}P_{2,i},\,\,
\rho=\frac{\frac{1}{N}\sum_{i=1}^{N}E[X_{1,i}X_{2,i}]}{\sqrt{P_{1}P_{2}}}.\nonumber\\
\end{eqnarray}
Letting $\epsilon\rightarrow 0$,
$R_{1}+R_{2}\leq \frac{1}{2}\log(1+\frac{P_{1}+P_{2}+2\sqrt{P_{1}P_{2}}\rho}{\sigma_{1}^{2}})
-\frac{1}{2}\log(1+\frac{P_{1}+P_{2}+2\sqrt{P_{1}P_{2}}\rho}{\sigma_{1}^{2}+\sigma_{2}^{2}})$ is proved.

Now it remains to show that $R_{2}\leq\frac{1}{2}\log(1+\frac{(1-\rho^{2})P_{2}}{\sigma_{1}^{2}})$, and the proof is exactly in the same fashion as that
in \cite[pp. 627-628]{cover7}. Hence we omit the proof here.
The proof of Corollary \ref{T6} is completed.

\section{Proof of Corollary \ref{T9}\label{appen2}}

We begin with the sum rate bound $R_{1}+R_{2}-\frac{1}{N}H(W_{1},W_{2}|Z^{N})\leq \epsilon$, which can be bounded by
\begin{eqnarray}\label{xxs.sy6}
&&R_{1}+R_{2}-\epsilon\leq \frac{1}{N}H(W_{1},W_{2}|Z^{N})\nonumber\\
&&\stackrel{(a)}\leq \frac{1}{N}(I(W_{1},W_{2};Y^{N}|Z^{N})+\delta(\epsilon))\nonumber\\
&&\stackrel{(b)}\leq \frac{1}{N}(I(X_{1}^{N},X_{2}^{N},S^{N};Y^{N}|Z^{N})+\delta(\epsilon))\nonumber\\
&&\stackrel{(c)}=\frac{1}{N}(I(X_{1}^{N},X_{2}^{N},S^{N};Y^{N})-I(X_{1}^{N},X_{2}^{N},S^{N};Z^{N})+\delta(\epsilon))\nonumber\\
&&=\frac{1}{N}\sum_{i=1}^{N}(H(Y_{i}|Y^{i-1})-H(Y_{i}|X_{1,i},X_{2,i},S_{i})-H(Z_{i}|Z^{i-1})+H(Z_{i}|X_{1,i},X_{2,i},S_{i}))+\frac{\delta(\epsilon)}{N}\nonumber\\
&&\stackrel{(d)}=\frac{1}{N}\sum_{i=1}^{N}(H(Y_{i}|Y^{i-1},Z^{i-1})-H(Y_{i}|X_{1,i},X_{2,i},S_{i})-H(Z_{i}|Z^{i-1})+H(Z_{i}|X_{1,i},X_{2,i},S_{i}))+\frac{\delta(\epsilon)}{N}\nonumber\\
&&\leq \frac{1}{N}\sum_{i=1}^{N}(H(Y_{i}|Z^{i-1})-H(Y_{i}|X_{1,i},X_{2,i},S_{i})-H(Z_{i}|Z^{i-1})+H(Z_{i}|X_{1,i},X_{2,i},S_{i}))+\frac{\delta(\epsilon)}{N}\nonumber\\
&&\stackrel{(e)}\leq \frac{1}{N}\sum_{i=1}^{N}(H(Y_{i})-H(Y_{i}|X_{1,i},X_{2,i},S_{i})-H(Z_{i})+H(Z_{i}|X_{1,i},X_{2,i},S_{i}))+\frac{\delta(\epsilon)}{N}\nonumber\\
&&=\frac{1}{N}\sum_{i=1}^{N}(I(X_{1,i},X_{2,i},S_{i};Y_{i})-I(X_{1,i},X_{2,i},S_{i};Z_{i}))+\frac{\delta(\epsilon)}{N},
\end{eqnarray}
where (a) follows from Fano's inequality and $P_{e}\leq \epsilon$, (b) follows from $H(W_{1},W_{2}|X_{1}^{N},X_{2}^{N})=0$,
(c) follows from
$(X_{1}^{N},X_{2}^{N},S^{N})\rightarrow Y^{N}\rightarrow Z^{N}$, (d) follows from $Y_{i}\rightarrow Y^{i-1}\rightarrow Z^{i-1}$, and
(e) follows from $Z^{i-1}\rightarrow Y_{i}\rightarrow Z_{i}$, which indicates that $I(Z_{i};Z^{i-1})\leq I(Y_{i};Z^{i-1})$, i.e.,
$H(Y_{i}|Z^{i-1})-H(Z_{i}|Z^{i-1})\leq H(Y_{i})-H(Z_{i})$.

Then substituting $Y_{i}=X_{1,i}+X_{2,i}+S_{i}+\eta_{1,i}$ and $Z_{i}=Y_{i}+\eta_{2,i}$ into (\ref{xxs.sy6}),
we have
\begin{eqnarray}\label{xxs.xiaobao3}
&&R_{1}+R_{2}-\epsilon\nonumber\\
&&\leq\frac{1}{N}\sum_{i=1}^{N}h(Y_{i})-\frac{1}{N}\sum_{i=1}^{N}h(Y_{i}|X_{1,i},X_{2,i},S_{i})
-\frac{1}{N}\sum_{i=1}^{N}h(Z_{i})+\frac{1}{N}\sum_{i=1}^{N}h(Z_{i}|X_{1,i},X_{2,i},S_{i})+\frac{\delta(\epsilon)}{N}\nonumber\\
&&\stackrel{(f)}=\frac{1}{N}\sum_{i=1}^{N}h(Y_{i})-\frac{1}{N}\sum_{i=1}^{N}h(\eta_{1,i})
-\frac{1}{N}\sum_{i=1}^{N}h(Z_{i})+\frac{1}{N}\sum_{i=1}^{N}h(\eta_{1,i}+\eta_{2,i})+\frac{\delta(\epsilon)}{N}\nonumber\\
&&\stackrel{(g)}\leq\frac{1}{N}\sum_{i=1}^{N}h(Y_{i})-\frac{1}{2}\log2\pi e\sigma_{1}^{2}
-\frac{1}{N}\sum_{i=1}^{N}\frac{1}{2}\log(2^{2h(Y_{i})}+2^{2h(\eta_{2,i})})+\frac{1}{2}\log2\pi e(\sigma_{1}^{2}+\sigma_{2}^{2})+\frac{\delta(\epsilon)}{N}\nonumber\\
&&\stackrel{(h)}\leq\frac{1}{N}\sum_{i=1}^{N}h(Y_{i})-\frac{1}{2}\log2\pi e\sigma_{1}^{2}
-\frac{1}{2}\log(2^{2\frac{1}{N}\sum_{i=1}^{N}h(Y_{i})}+2\pi e\sigma_{2}^{2})+\frac{1}{2}\log2\pi e(\sigma_{1}^{2}+\sigma_{2}^{2})+\frac{\delta(\epsilon)}{N}\nonumber\\
&&\stackrel{(i)}\leq\frac{1}{2}\log2\pi e(P_{1}+P_{2}+2\sqrt{P_{1}P_{2}}\rho+\sigma_{1}^{2})-\frac{1}{2}\log2\pi e\sigma_{1}^{2}
-\frac{1}{2}\log(2\pi e(P_{1}+P_{2}+2\sqrt{P_{1}P_{2}}\rho+\sigma_{1}^{2}+\sigma_{2}^{2}))\nonumber\\
&&+\frac{1}{2}\log2\pi e(\sigma_{1}^{2}+\sigma_{2}^{2})+\frac{\delta(\epsilon)}{N}\nonumber\\
&&=\frac{1}{2}\log(1+\frac{P_{1}+P_{2}+2\sqrt{P_{1}P_{2}}\rho}{\sigma_{1}^{2}})
-\frac{1}{2}\log(1+\frac{P_{1}+P_{2}+2\sqrt{P_{1}P_{2}}\rho}{\sigma_{1}^{2}+\sigma_{2}^{2}})+\frac{\delta(\epsilon)}{N},
\end{eqnarray}
where (f) follows from $\eta_{1,i}$ and $\eta_{2,i}$ are independent of $X_{1,i}$ and $X_{2,i}$, (g) follows from
the entropy power inequality, (h) follows from the fact that $\log(2^{x}+c)$ is a convex function and Jensen's inequality,
and (i) follows from $\frac{1}{N}\sum_{i=1}^{N}h(Y_{i})-\frac{1}{2}\log(2^{2\frac{1}{N}\sum_{i=1}^{N}h(Y_{i})}+2\pi e\sigma_{2}^{2})$
is increasing while $\frac{1}{N}\sum_{i=1}^{N}h(Y_{i})$ is increasing and
\begin{eqnarray}\label{xxs.xiaobao4}
&&\frac{1}{N}\sum_{i=1}^{N}h(Y_{i})\nonumber\\
&&\leq \frac{1}{N}\sum_{i=1}^{N}\frac{1}{2}\log2\pi e(P_{1,i}+P_{2,i}+Q
+2E[X_{1,i}X_{2,i}]+2E[X_{1,i}S_{i}]+2E[X_{2,i}S_{i}]+\sigma_{1}^{2})\nonumber\\
&&\leq\frac{1}{2}\log2\pi e(\frac{1}{N}\sum_{i=1}^{N}(P_{1,i}+P_{2,i}+2E[X_{1,i}X_{2,i}]+2E[X_{1,i}S_{i}]+2E[X_{2,i}S_{i}])+\sigma_{1}^{2})\nonumber\\
&&\stackrel{(j)}=\frac{1}{2}\log2\pi e(P_{1}+P_{2}+2\sqrt{P_{1}P_{2}}\rho_{12}+2\sqrt{P_{1}Q}\rho_{1s}+2\sqrt{P_{2}Q}\rho_{2s}+\sigma_{1}^{2}),
\end{eqnarray}
where (j) follows from the definitions
\begin{eqnarray}\label{xxs.xiaobao2}
&&E[X^{2}_{1,i}]=P_{1,i},\,\,E[X^{2}_{2,i}]=P_{2,i},\,\,P_{1}=\frac{1}{N}\sum_{i=1}^{N}P_{1,i},\,\,P_{2}=\frac{1}{N}\sum_{i=1}^{N}P_{2,i},\nonumber\\
&&\rho_{12}=\frac{\frac{1}{N}\sum_{i=1}^{N}E[X_{1,i}X_{2,i}]}{\sqrt{P_{1}P_{2}}}\,\,
\rho_{1s}=\frac{\frac{1}{N}\sum_{i=1}^{N}E[X_{1,i}S_{i}]}{\sqrt{P_{1}Q}},\,\,\rho_{2s}=\frac{\frac{1}{N}\sum_{i=1}^{N}E[X_{2,i}S_{i}]}{\sqrt{P_{2}Q}}.\nonumber\\
\end{eqnarray}
Letting $\epsilon\rightarrow 0$,
the sum rate bound of Theorem \ref{T9} is proved.

Now it remains to show the upper bound on the individual rate $R_{2}$, see the details below.
First, note that
\begin{eqnarray}\label{gg1.xxs.sy6}
&&R_{2}-\epsilon\leq \frac{1}{N}H(W_{2}|Z^{N})\leq \frac{1}{N}H(W_{2})\stackrel{(k)}=\frac{1}{N}H(W_{2}|X_{1}^{N},S^{N})\nonumber\\
&&\stackrel{(l)}\leq \frac{1}{N}(I(W_{2};Y^{N}|X_{1}^{N},S^{N})+\delta(\epsilon))\nonumber\\
&&\stackrel{(m)}\leq \frac{1}{N}(I(X_{2}^{N};Y^{N}|X_{1}^{N},S^{N})+\delta(\epsilon))\nonumber\\
&&\leq\frac{1}{N}\sum_{i=1}^{N}(H(Y_{i}|X_{1,i},S_{i})-H(Y_{i}|X_{1,i},X_{2,i},S_{i}))+\frac{\delta(\epsilon)}{N},
\end{eqnarray}
where (k) follows from $W_{2}$ is independent of $X_{1}^{N}$ and $S^{N}$, (l) follows from Fano's inequality and $P_{e}\leq \epsilon$,
and (m) follows from $H(W_{2}|X_{2}^{N})=0$.

Then substituting $Y_{i}=X_{1,i}+X_{2,i}+S_{i}+\eta_{1,i}$ into (\ref{gg1.xxs.sy6}), and using the fact that $\eta_{1,i}$
is independent of $X_{1,i}$, $X_{2,i}$ and $S_{i}$,
we have
\begin{eqnarray}\label{gg1.xxs.xiaobao3}
&&R_{2}-\epsilon\leq \frac{1}{N}\sum_{i=1}^{N}(h(X_{2,i}+\eta_{1,i}|X_{1,i},S_{i})-h(\eta_{1,i}))+\frac{\delta(\epsilon)}{N}\nonumber\\
&&\stackrel{(n)}\leq\frac{1}{N}\sum_{i=1}^{N}(\frac{1}{2}\log2\pi e(Var(X_{2,i}|X_{1,i},S_{i})+\sigma_{1}^{2})-\frac{1}{2}\log2\pi e\sigma_{1}^{2})+\frac{\delta(\epsilon)}{N}\nonumber\\
&&\stackrel{(o)}\leq\frac{1}{N}\sum_{i=1}^{N}(\frac{1}{2}\log2\pi e(Var(X_{2,i}-a_{i}X_{1,i}-b_{i}S_{i})+\sigma_{1}^{2})-\frac{1}{2}\log2\pi e\sigma_{1}^{2})+\frac{\delta(\epsilon)}{N}\nonumber\\
&&=\frac{1}{N}\sum_{i=1}^{N}(\frac{1}{2}\log2\pi e(P_{2,i}+a_{i}^{2}P_{1,i}+b_{i}^{2}Q-2a_{i}E[X_{1,i}X_{2,i}]-2b_{i}E[X_{2,i}S_{i}]+2a_{i}b_{i}E[X_{1,i}S_{i}])
+\sigma_{1}^{2})\nonumber\\
&&-\frac{1}{2}\log2\pi e\sigma_{1}^{2})+\frac{\delta(\epsilon)}{N},
\end{eqnarray}
where (n) follows from $\eta_{1,i}$ is independent of $X_{1,i}$, $X_{2,i}$ and $S_{i}$, and (o) follows from
$Var(X_{2,i}|X_{1,i},S_{i})$ is no greater than the variance of the difference between $X_{2,i}$ and its linear MMSE estimation
$\hat{X}_{2,i}=a_{i}X_{1,i}+b_{i}S_{i}$, and
\begin{eqnarray}\label{guiguan1}
&&a_{i}=\frac{E[X_{1,i}X_{2,i}]Q-E[X_{1,i}S_{i}]E[X_{2,i}S_{i}]}{P_{1,i}Q-(E[X_{1,i}S_{i}])^{2}},\,\,
b_{i}=\frac{E[X_{2,i}S_{i}]P_{1,i}-E[X_{1,i}S_{i}]E[X_{1,i}X_{2,i}]}{P_{1,i}Q-(E[X_{1,i}S_{i}])^{2}}.\nonumber\\
\end{eqnarray}

Observing that in (\ref{gg1.xxs.xiaobao3}), we can readily check that the logarithm function is concave in $P_{1,i}$, $P_{2,i}$,
$E[X_{1,i}X_{2,i}]$, $E[X_{2,i}S_{i}]$ and $E[X_{1,i}S_{i}]$ by evaluating the corresponding Hessian matrix.
Hence applying Jensen's inequality, using (\ref{xxs.xiaobao2}), defining
\begin{eqnarray}\label{guiguan2}
a=\sqrt{\frac{P_{2}}{P_{1}}}\frac{\rho_{12}-\rho_{1s}\rho_{2s}}{1-\rho^{2}_{1s}},\,\,\,b=\sqrt{\frac{P_{2}}{Q}}\frac{\rho_{2s}-\rho_{12}\rho_{1s}}{1-\rho^{2}_{1s}},
\end{eqnarray}
and letting $\epsilon\rightarrow 0$,
the bound on the individual rate $R_{2}$ is proved.

The proof of Corollary \ref{T9} is completed.


\begin{thebibliography}{99}


\bibitem{mac2} H. D. Liao, ``Multiple-access channels,'' Ph.D. dissertation, Univ.
Hawaii, Honolulu, 1972.

\bibitem{gmac1} T. Cover, ``Some advances in broadcast channels,'' in {\sl Advances in
Communication Systems}, vol. 4, A. Viterbi, Ed. San Francisco:
Academic Press, 1975.


\bibitem{cover2} T. M. Cover and C. S. K. Leung, ``An achievable rate region for the
multiple-access channel with feedback,'' {\sl IEEE Trans. Inf. Theory}, vol. 27, no. 3, pp. 292-298, 1981.


\bibitem{cover4} R. Venkataramanan and S. S. Pradhan, ``A new achievable rate region for the
multiple-access channel with noiseless feedback,'' {\sl IEEE Trans. Inf. Theory}, vol. 57, no. 12, pp. 8038-8054, 2011.


\bibitem{cover7} L. H. Ozarow, ``The capacity of the white Gaussian multiple access channel with feedback,'' {\sl IEEE Trans. Inf. Theory}, vol.
27, no. 5, pp. 292-298, 1981.

\bibitem{sk} J. P. M. Schalkwijk and T. Kailath, ``A coding scheme
for additive noise channels with feedback. part I: No bandwidth constraint,'' {\sl IEEE Trans. Inf. Theory}, vol. 12, pp. 172-182, 1966.

\bibitem{cover4.x} A. Rosenzweig, ``The capacity of Gaussian multi-user channels with state and feedback,'' {\sl IEEE Trans. Inf. Theory}, vol.
53, no. 11, pp. 4349-4355, 2007.

\bibitem{dms1} A. Bracher, and A. Lapidoth,
``Feedback, cribbing, and causal state information on the multiple-access channel,''
{\sl IEEE Trans. Inf. Theory}, vol. 60, no. 12, pp. 7627-7654, 2014.

\bibitem{dms2} O. Sabag, H. H. Permuter and S. Shamai, ``Capacity-achieving coding scheme for the MAC
with degraded message sets and feedback,'' {\sl 2019 IEEE International Symposium on Information Theory (ISIT)}, pp. 2259-2263, 2019.

\bibitem{pm} O. Shayevitz and M. Feder, ``A simple proof for the optimality of randomized posterior matching,''
{\sl IEEE Trans. Inf. Theory}, vol. 62, no. 6, pp. 3410-3418, 2016. 

\bibitem{Wy} A. D. Wyner, ``The wire-tap channel,''
{\sl Bell Syst. Tech. J.}, vol. 54, no. 8, pp.
1355-1387, 1975.

\bibitem{CK} I. Csisz$\acute{a}$r and J. K\"{o}rner, ``Broadcast channels with confidential messages,'' {\sl IEEE Trans.
Inf. Theory}, vol. 24, no. 3, pp. 339-348, 1978.

\bibitem{hellman} S. K. Leung-Yan-Cheong and M. E. Hellman,
``The Gaussian wire-tap channel,'' {\sl IEEE Trans. Inf. Theory}, vol. 24, no. 4, pp. 451-456, 1978.

\bibitem{mactifs1} E. Tekin and A. Yener, ``The Gaussian multiple access wire-tap channel,''
{\sl IEEE Trans. Inf. Theory}, vol. 54, no. 12, pp. 5747-5755, 2008.

\bibitem{mactifs2} Y. Chen, D. He and Y. Luo, ``Strong secrecy of arbitrarily varying
multiple access channels,'' {\sl IEEE Trans. Inf. Forensics and Security}, vol. 16, no. 7, pp. 3662-3677, 2021.

\bibitem{mactifs3} R. Fritschek and G. Wunder, ``On the Gaussian multiple access wiretap channel
and the Gaussian wiretap channel with a helper:
achievable schemes and upper bounds,'' {\sl IEEE Trans. Inf. Forensics and Security}, vol. 14, no. 5, pp. 1224-1239, 2019.

\bibitem{mactifs4} H. ZivariFard, M. R. Bloch and A. Nosratinia, ``Two-multicast channel with confidential messages,'' {\sl IEEE Trans. Inf. Forensics and Security}, 
vol. 16, no. 4, pp. 2743-2758, 2021.


\bibitem{mactifs5} P. Xu, Z. Ding, X. Dai, and K. K. Leung, ``Rate regions for multiple access channel with
conference and secrecy constraints,''
{\sl IEEE Trans. Inf. Forensics and Security}, vol. 8, no. 12,
pp. 1961-1974, 2013.

\bibitem{mactifs6} H. He, X. Luo, J. Weng and K. Wei, ``Secure transmission in multiple access wiretap
channel: cooperative jamming
without sharing CSI,'' {\sl IEEE Trans. Inf. Forensics and Security},
vol. 16, no. 6, pp. 3401-3411, 2021.

\bibitem{mactifs7} A. Sonee and G. A. Hodtani, ``On the secrecy rate region of multiple-access wiretap channel with noncausal side information,''
{\sl IEEE Trans. Inf. Forensics and Security}, vol. 10, no. 6, pp. 1151-1166, 2015.

\bibitem{mactifs8} B. Dai, Z. Ma, M. Xiao, X. Tang and P. Fan, ``Secure communication over finite state multiple-access wiretap channel with delayed feedback,''
{\sl IEEE Journal on Selected Areas in Communications}, vol. 36, no. 4, pp. 723-736, 2018.

\bibitem{mactifs9} B. Dai and Z. Ma, ``Multiple-access relay wiretap channel,''
{\sl IEEE Trans. Inf. Forensics and Security}, vol. 10, no. 9, pp. 1835-1849, 2015.


\bibitem{gunx} D. Gunduz, D. R. Brown and H. V. Poor, ``Secret communication with feedback,'' {\sl International Symposium on
Information Theory and Its Applications, ISITA 2008}, pp. 1-6, 2008.


\bibitem{lichong} C. Li, Y. Liang, H. V. Poor and S. Shamai, ``Secrecy capacity of colored Gaussian noise channels with feedback,''
{\sl IEEE Trans. Inf. Theory}, vol. 65, no. 9, pp. 5771-5782, 2019.


\bibitem{daitit} B. Dai, C. Li, Y. Liang, Z. Ma and S. Shamai, ``Impact of action-dependent state and channel feedback on Gaussian wiretap channels,''
{\sl IEEE Trans. Inf. Theory}, vol. 66, no. 6, pp. 3435-3455, 2020.

\bibitem{daiisita} B. Dai, C. Li, Y. Liang, Z. Ma and S. Shamai, ``On the capacity of Gaussian multiple-access wiretap channels with feedback,''
{\sl International Symposium on
Information Theory and Its Applications, ISITA 2020}, pp. 397-401, 2020.


\bibitem{mac-dms} D. Slepian and J. K. Wolf, ``A coding theorem for multiple access
channels with correlated sources,'' {\sl Bell Syst. Tech. J.}, vol. 51, no. 7,
pp. 1037-1076, 1973.

\bibitem{kim1} Y. Kim, A. Sutivong and S. Sigurj$\acute{o}$nsson ``Multiple user writing on dirty paper,''
{\sl 2004 IEEE International Symposium on Information Theory (ISIT)}, p. 534, 2004.

\bibitem{tangxiaojun} X. Tang, R. Liu, P. Spasojevi$\acute{c}$ and H. V. Poor, ``Multiple access
channels with
generalized feedback and confidential messages,'' {\sl 2007 IEEE Information Theory
Workshop (ITW)}, pp. 1-6, 2004.

\bibitem{c3} R. G. Gallager and B. Nakiboglu,
``Variations on a theme by Schalkwijk and Kailath,''
{\sl IEEE Trans. Inf. Theory}, vol. 56, no. 1, pp. 6-17, 2010.


\end{thebibliography}
\end{document}